\documentclass[iicol, referee, pdflatex, sn-mathphys-num]{sn-jnl}

\usepackage{graphicx}%
\usepackage{multirow}%
\usepackage{amsmath,amssymb,amsfonts}%
\usepackage{amsthm}%
\usepackage{mathrsfs}%
\usepackage[title]{appendix}%
\usepackage{xcolor}%
\usepackage{textcomp}%
\usepackage{manyfoot}%
\usepackage{booktabs}%
\usepackage{algorithm}%
\usepackage{algorithmicx}%
\usepackage{algpseudocode}%
\usepackage{listings}%

\usepackage{siunitx} 
\usepackage{upgreek} 
\usepackage{bm} 
\usepackage{tabularx}
\onecolumn 

\theoremstyle{thmstyleone}%
\theoremstyle{thmstyletwo}%

\theoremstyle{thmstylethree}%

\def\paperTitle{
Reconfigurable generation of high-power structured light via nonlinear beam shaping 
}
\begin{document}
\title[High-power structured light via nonlinear beam shaping]{\paperTitle}

\author[1]{
    \fnm{KyeoReh}
    \sur{Lee}}
\email{kyeoreh.lee@yale.edu}

\author[1]{
    \fnm{Baichuan}
    \sur{Huang}}
\email{baichuan.huang@yale.edu}

\author[1,2]{
    \fnm{Peyman}
    \sur{Ahmadi}}
\email{peyman.ahmadi@scilase.com}

\author[1]{
    \fnm{Mert}
    \sur{Ercan}}
\email{mert.ercan@yale.edu}

\author[1,3]{
    \fnm{Stefan}
    \sur{Rothe}}
\email{stefan.rothe@utwente.nl}

\author[1,4]{
    \fnm{Chun-Wei}
    \sur{Chen}}
\email{cwc218@bath.ac.uk}

\author*[1]{
    \fnm{Hui}
    \sur{Cao}}
\email{hui.cao@yale.edu}

\affil[1]{
    \orgdiv{Department of Applied Physics},
    \orgname{Yale University},
    \orgaddress{\city{New Haven}, \postcode{06520}, \state{Connecticut}, \country{USA}}}
\affil[2]{
    \orgname{Sci Lase},
    \orgaddress{\city{Sheridan}, \postcode{82801}, \state{Wyoming}, \country{USA}}}
\affil[3]{
    \orgdiv{MESA+ Institute for Nanotechnology},
    \orgname{University of Twente},
    \orgaddress{\city{Enschede}, \postcode{7500 AE}, \country{The Netherlands}}}
\affil[4]{
    \orgdiv{Centre for Photonics, Department of Physics},
    \orgname{University of Bath},
    \orgaddress{\city{Bath}, \postcode{BA2 7AY}, \country{UK}}}

\abstract{
    High-power structured light has a wide range of applications, from material processing and high-capacity optical communications to programmable electron beams, plasmas, and nuclear states. On-demand generation of structured light and adaptive control of beam profiles are essential for many practical applications, but are difficult to achieve at high power. Here, we demonstrate reconfigurable generation of structured light from a high-power multimode-fiber laser amplifier through full-field control of a low-power seed. An efficient nonlinear beam shaping scheme based on a local linear approximation of the complex nonlinear input--output relation is developed and realized \textit{in situ}. Diverse structured beams such as Gaussian, Bessel, vector and orbital angular momentum beams are directly generated from the fiber amplifier at powers exceeding 500 W. Our scheme enables real-time programmability of structured light and is readily scalable to even higher power levels. This work provides a general framework for controlling high-dimensional nonlinear systems without accurate knowledge or tractable model.
}
\maketitle

\section*{Introduction}\label{sec:Introduction}

\begin{figure}[ht]
\centering
\includegraphics[width=0.7\textwidth]{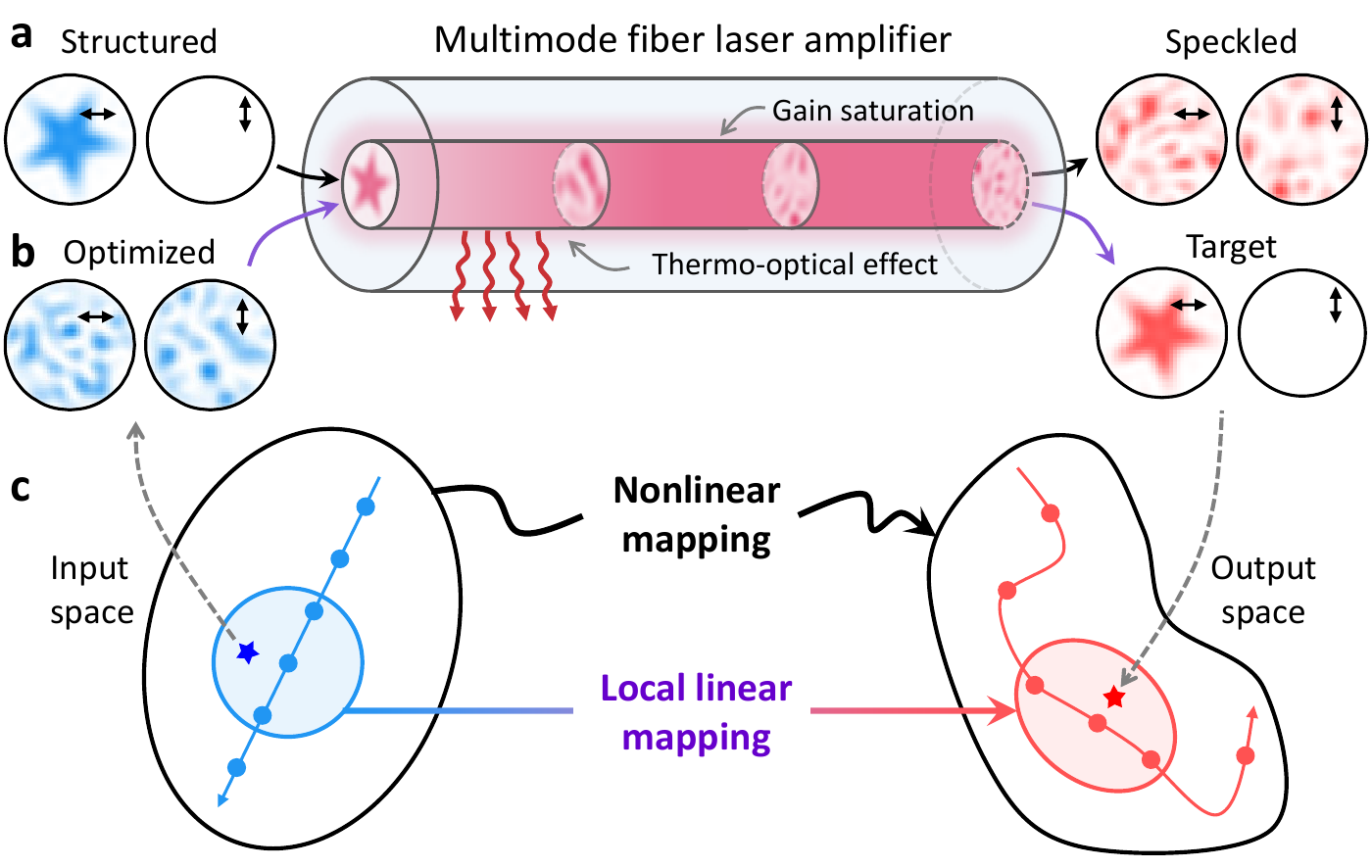}
\caption{
    \textbf{High-power structured light generation via an MMF amplifier.}
    Schematic of a multimode fiber (MMF) laser amplifier with gain saturation and thermo-optical nonlinearity. 
    \textbf{a}, An input structured beam (left) experiences linear and nonlinear mode coupling and polarization mixing in the MMF amplifier, producing a depolarized speckle output (right). The black arrows in the upper-right corners of the beam profiles indicate the linear polarization directions.
    \textbf{b}, The optimized input vector field (solution) yielding the target structured output beam (target).
    \textbf{c}, Conceptual illustration of the nonlinear input--output mapping and the local transmission matrix (TM) framework. A straight line in the input space (blue solid line) is transformed into a complex contour in the output space (red solid line) by the nonlinear mapping of an MMF amplifier. The local linear regression approximates the mapping between local input and output regions (shaded areas). The input solution and target output are represented as stars in the input and output spaces, respectively.
}\label{fig:1Overview}
\end{figure}

Structured beams have been generated by placing beam-shaping elements inside or outside laser cavities \cite{forbes2021structured, harrison2024progress, carbajo2025structured}. The power-handling capability of beam-shaping devices, exacerbated by the intra-cavity power enhancement factor, limits the output power of structured light lasers \cite{forbes2019structured}. Extra-cavity beam shaping can produce higher-power structured light \cite{harrison2024progress, carbajo2025structured}, but reconfigurable beam shapers typically cannot sustain optical powers exceeding \qty{1}{\kW}. A straightforward solution would be to create low-power structured light and subsequently amplify it \cite{piehler2013amplification, loescher2015radially, lin2018106, sroor2018purity, liu2024all}. However, amplifying structured light without distortion is extremely difficult in highly multimoded, high-power amplifiers due to mode-dependent gain, thermo-optical effects, gain saturation, and nonlinear mode coupling. A multimode fiber (MMF) laser amplifier is a representative example (Fig.~\ref{fig:1Overview}a). While MMF amplifiers could provide significantly higher gain than solid-state and semiconductor optical amplifiers, their output beam quality would deteriorate severely due to strong linear and nonlinear mode coupling and polarization mixing \cite{florentin2017shaping, florentin2019shaping, rothe2025output, rothe2025wavefront}. Thus, most realizations of structured light amplification are based on few-mode, multi-core, or ring-core fibers that minimize such intermodal coupling effects \cite{montoya2017photonic, wang2018transverse, lin2020reconfigurable, lin2021high, ou2022amplification, ji2023controlled, liu2024high, zalesskaia2026high}.

Here we demonstrate the direct generation of high-power structured light from a highly multimode fiber laser amplifier through low-power input beam shaping (Fig.~\ref{fig:1Overview}b). The key challenge is finding the input beam shape that precompensates for all complex coupling effects. Since the input--output relation of a high-power MMF amplifier is highly nonlinear, theoretical or numerical modeling is intractable, and existing beam-shaping methods for linear or low-order nonlinear systems become invalid \cite{strudley2014ultrafast, florentin2019shaping, fleming2019perturbation, moon2023measuring, ni2023nonlinear, gutierrez2024reaching}. Two broad classes of experimental approaches have been explored: gradient-free optimization and data-driven neural-network prediction. Both have intrinsic limitations. \textit{In situ} optimization of inputs using gradient-free methods, including random search \cite{nasiri2014adaptive, florentin2017shaping, florentin2018space, deliancourt2019wavefront, cui2025optical, yanagimoto2025programmable, rothe2025wavefront}, genetic algorithms \cite{tzang2018adaptive, fleming2019nonlinear, wei2020harnessing, zhang2022spectrally, hary2023tailored}, particle swarm optimization \cite{hary2025spectral}, and simulated annealing \cite{finkelstein2023spectral}, does not guarantee convergence to the optimal solution. In our previous work, for example, such algorithms failed to reach high beam quality and polarization purity \cite{rothe2025wavefront}. Trained neural networks can learn an approximate input--output relation \cite{teugin2020controlling, yan2024image}, but they become unreliable for output beam shapes that are rare or absent in the training set, which is precisely the regime most relevant to on-demand structured light generation.

To overcome these limitations, we introduce a robust and efficient nonlinear beam shaping method that addresses strong nonlinearity in a multimode system. It is based on a linear approximation of the nonlinear input--output function within a local region of the input space. As this region progressively contracts, the linear approximation becomes increasingly accurate and converges toward the optimal solution. Applying this framework to a high-power single-frequency MMF amplifier, we generate various structured beams with high precision at an output power of \qty{538}{\W}.

By modulating both amplitude and phase of the input field in two orthogonal polarizations, we obtain an output beam with beam propagation factor $M^2 = \num{1.09}$ and a polarization extinction ratio (PER) of \qty{19.6}{\dB}. Such high beam quality and polarization purity address the major concern for MMF amplifiers, which typically produce speckled, depolarized outputs. Our method also enables direct generation of complex patterns from high-power MMF amplifiers for diverse applications. Examples demonstrated here include multiple focused spots for parallel machining, a bright central spot with side lobes to facilitate material welding, a doughnut beam with tunable orbital angular momentum (OAM) for particle manipulation, a Bessel beam with extended depth of focus for volumetric processing, and a radially polarized vector beam for tight focusing. Our approach allows on-the-fly correction of the output beam profile through real-time adjustment of the input to the amplifier. It avoids high-power handling and enables further power scaling of structured light generation.

More broadly, this work addresses the fundamental challenge of controlling complex nonlinear systems by offering a general and reliable control strategy based on local linearization. It can be extended beyond multimode fiber amplifiers to a wide range of high-dimensional nonlinear systems, where neither model-based approaches nor data-driven methods alone provide optimal solutions. 

\section*{Results}\label{sec:Results}

\subsection*{Local transmission matrix}\label{subsec:LocalTM}

In linear systems, the input--output mapping is fully described by a transmission matrix (TM). Although it remains applicable to a dissipative system with linear gain or loss, the TM fails in a nonlinear system such as an MMF amplifier as the superposition principle no longer holds \cite{fleming2019perturbation, goicoechea2025detecting}. This is illustrated in Fig.~\ref{fig:1Overview}c, where a straight line in the input space is transformed into a complex contour in the output space. Our goal is to find the input (solution) that produces a desired output (target) represented by the stars in Fig.~\ref{fig:1Overview}c. 

Although the complex and nonlinear input--output mapping of the MMF amplifier cannot be modeled analytically, it can still be estimated experimentally. We measure the TM of the system as in linear systems by exploring the full input space (Fig.~\ref{fig:1Overview}c). The resulting \textit{global} TM provides a linear approximation of the nonlinear mapping over the entire input space (domain). We then conduct gradient-based optimization using the gradient derived from the global TM. Since the gradient is inaccurate, the optimization does not converge to the solution, although it may get closer. Here, the central point is that the nonlinear input--output mapping of the MMF amplifier becomes increasingly linear within a smaller domain. We exploit this by re-measuring the TM within a contracted local region around the best estimate from the previous optimization and denote this linear approximation as a \textit{local} TM (Fig.~\ref{fig:1Overview}c). Within this contracted domain, the local TM captures the gradient more accurately, restoring the power of gradient-based optimization. By iterating this process of contracting the domain and refining the local TM, the optimization converges reliably to the solution with high precision, without prior knowledge of the nonlinear mapping. We confirm the validity of the local TM framework through numerical simulations of the MMF amplifier (see Methods, Fig.~\ref{figS:simulation}). Detailed procedures are provided in the following sections, together with the experimental implementation and measured results.

\subsection*{Multimode fiber amplifier}\label{subsec:MMFamplifier}
\begin{figure}[t]
    \centering
    \includegraphics[width=1.0\textwidth]{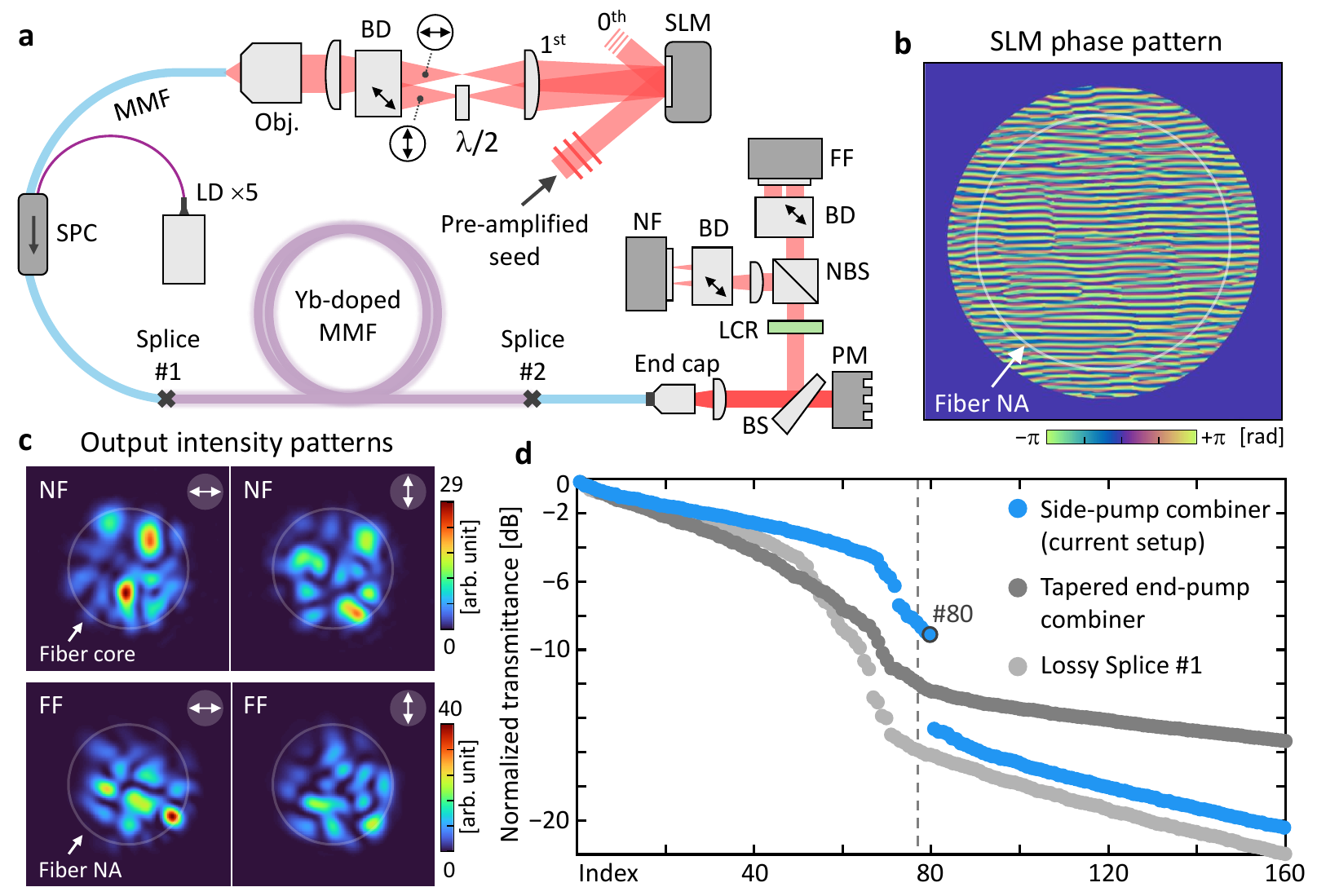}
    \caption{
        \textbf{MMF amplifier with input beam shaping.}
        \textbf{a}, Schematic of the MMF amplifier with input beam shaping and output characterization setup. See text for description. SLM, spatial light modulator; $\lambda/2$, half-wave plate; BD, beam displacer; Obj., objective lens; MMF, multimode fiber; SPC, signal-pump combiner; LD, laser diode; BS, Fresnel beam sampler; PM, power meter; LCR, liquid crystal retarder; NBS, non-polarizing beam splitter; NF and FF, near-field intensity distribution at the fiber output facet and far-field intensity distribution, $0^{\mathrm{th}}$ and $1^{\mathrm{st}}$, the zeroth-order and first-order diffractions from the SLM. Fiber splice points are marked by crosses.
        \textbf{b}, Phase-only SLM pattern. For visualization purposes, the linear phase ramp used for first-order diffraction and the SLM curvature-correction pattern are omitted.
        \textbf{c}, With unshaped input, the output NF and FF intensity distributions in horizontal and vertical linear polarizations at \qty{538}{\W}. The polarization direction is denoted by a white arrow at the top-right corner of each image.
        \textbf{d}, Normalized transmittance of the transmission eigenchannels in the MMF amplifier without pumping. The current setup with the side-pump combiner provides 80 transmission channels (blue dots), whereas the tapered end-pump combiner and a lossy splice result in a reduced number of transmission channels (dark and light gray dots, respectively). As a reference, the estimated number of transmission channels from the fiber V-number is 77 (vertical dashed line). Input and output patterns for representative transmission channels are presented in Fig.~\ref{figS:SingularVectors}.
    }\label{fig:Setup}
\end{figure}

The experimental setup is shown schematically in Fig.~\ref{fig:Setup}a; full details are provided in Fig.~\ref{figS:Setup}. Our MMF amplifier supports $N_\mathrm{fib}=80$ modes, corresponding to \num{40} spatial modes per polarization (see Methods). Continuous-wave output from a single-frequency laser operating at a wavelength of $\lambda_s = \qty{1064}{\nm}$ and power of \qty{100}{\mW} is pre-amplified to \qty{45}{\W} in a two-stage single-mode fiber amplifier and serves as the seed to the MMF amplifier. The signal--pump combiner (SPC) combines the seed with pump light from five laser diodes, each providing \qty{150}{\W} power at a wavelength of $\lambda_p = \qty{976}{\nm}$. A \num{4.8}-m-long Yb-doped MMF amplifies the signal to \qty{538}{\W}. The output power is limited by stimulated Brillouin scattering (SBS) \cite{rothe2025wavefront}. The amplifier is terminated with an anti-reflection coated end cap to prevent damage to the fiber facet. Both near-field (NF) and far-field (FF) intensity distributions for two orthogonal polarizations are simultaneously captured by two cameras equipped with beam displacers.

To achieve complete control over all spatial modes in both polarizations, we employ a vector-field control scheme for the input signal \cite{ploschner2015seeing, gomes2022near}. A liquid-crystal spatial light modulator (SLM), which modulates the phase of a single linear polarization (in the horizontal direction), is placed at the Fourier plane of the proximal facet of the MMF amplifier. The horizontally-polarized (H-pol) seed is incident onto the SLM, and its zeroth-order diffraction is suppressed by introducing a global phase ramp on the SLM. The two first-order diffraction beams are H-pol, and one of them passes through a half-wave plate to become vertically polarized (V-pol). A phase hologram is written to the SLM to encode distinct complex field distributions in the two beams with H- and V-pols. They are subsequently recombined by a beam displacer and enter an MMF amplifier (Fig.~\ref{fig:Setup}a). Due to the phase-only modulation capability of the liquid-crystal SLM, moderate power loss is unavoidable for full-field control. To minimize this loss for generating an arbitrary input vector field, we introduce a flat-top beam shaper and a dedicated algorithm for optimizing the SLM pattern (see Methods). The core-coupled seed power is \qty{8.2}{\W} and remains consistent across different vector input fields. Figure~\ref{fig:Setup}b shows an example of an optimized SLM phase pattern that has a grating-like structure to produce the two first orders.

Due to extensive mode coupling and polarization mixing within the MMF amplifier, the output beam is speckled and depolarized. 
A typical output beam profile without input optimization is shown in Fig.~\ref{fig:Setup}c. To determine the full output vector field, we acquire NF and FF images with linear (horizontal and vertical) as well as circular (left- and right-handed) polarizations by toggling the $\pi/2$ retardance of the liquid crystal retarder (LCR) (Fig.~\ref{fig:Setup}a). The spatial extents of NF and FF images are determined by the fiber core diameter and numerical aperture (NA), respectively.

Before input beam shaping, we confirm that all $N_\mathrm{fib}$ fiber modes survive throughout the MMF amplifier. This is crucial because if the effective number of controllable input modes (the degree of input control) were smaller than $N_\mathrm{fib}$, full output control would not be achievable even in passive MMF systems \cite{ploschner2015seeing}. Experimentally, we characterize the effective number of controllable modes by analyzing the TM of the MMF amplifier without pumping. The complete TM, encompassing both input and output spatial modes in two orthogonal polarizations, is retrieved from the measured NF and FF intensity patterns (see Methods). We confirm that the number of significant transmission eigenchannels equals $N_\mathrm{fib}$ (see Methods, Fig.~\ref{fig:Setup}d), proving that all $N_\mathrm{fib}$ fiber modes are controllable. We also find that the SPC design and fiber splice quality are critical factors determining the degree of control (Fig.~\ref{fig:Setup}d). Since an end-pump SPC usually has a tapered region, which reduces the effective number of propagating modes, we use a side-pump SPC so that all $N_\mathrm{fib}$ modes remain controllable. We validate the splice quality by comparing the number of significant transmission channels before and after each splice, confirming that both splice points in our MMF amplifier (marked by crosses in Fig.~\ref{fig:Setup}a) do not introduce notable loss of control.

\subsection*{\textit{In situ} optimization}\label{subsec:optimization}
\begin{figure}[t]
    \centering
    \includegraphics[width=0.9\textwidth]{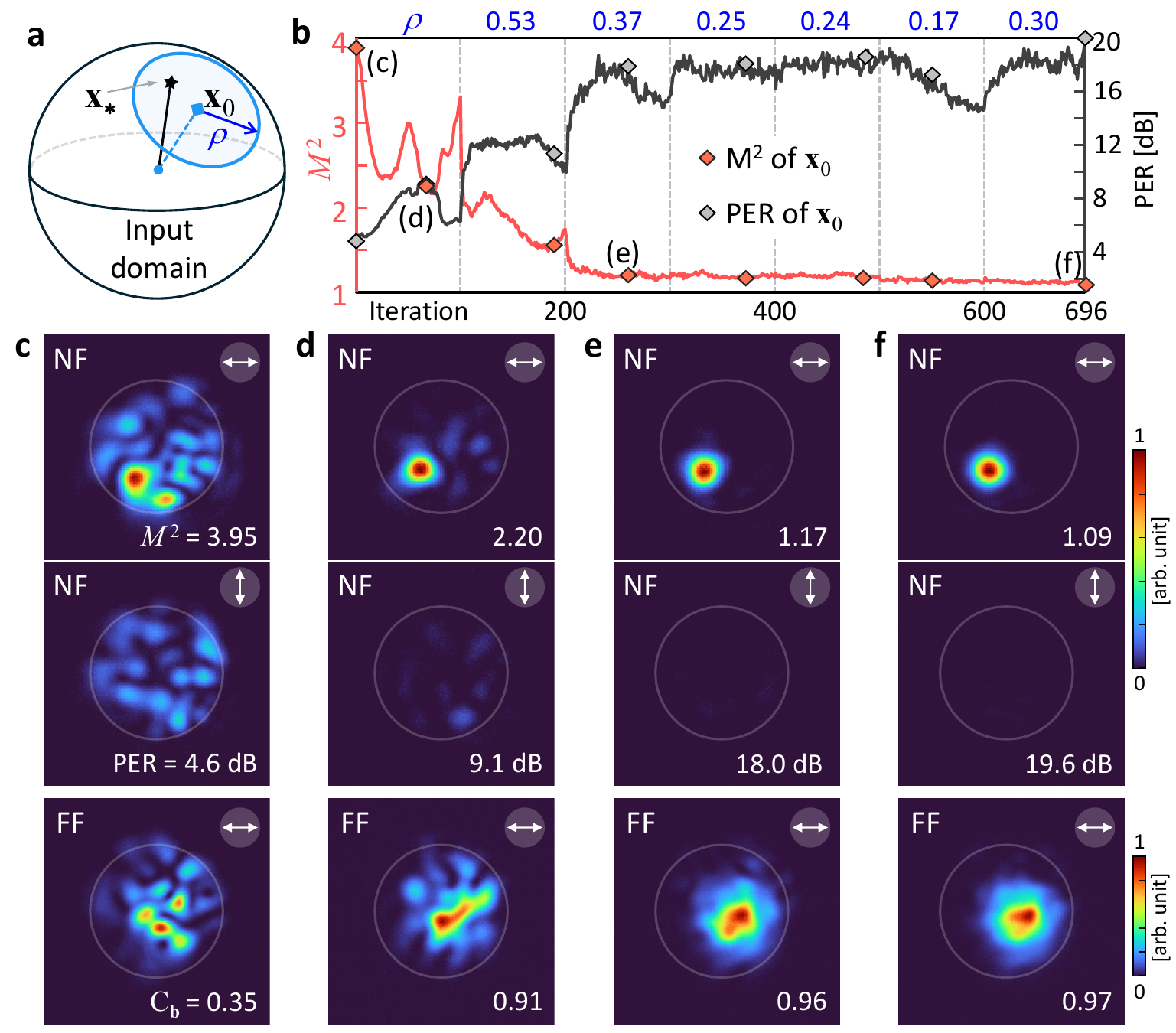}
    \caption{
        \textbf{Experimental demonstration of nonlinear beam shaping with local TM.}
        \textbf{a}, Illustration of the global and local domains. The local domain is centered at the reference vector $\mathbf{x}_0$ and has a size parameter $\rho$ to contain the solution ($\mathbf{x}_*$).
        \textbf{b}, \textit{In situ} convergence of the beam propagation factor ($M^2$, red) and polarization extinction ratio (PER, gray) over iterations. The gradient-based optimization consists of one global TM stage followed by six local TM stages with the corresponding $\rho$ values shown above the graph, each stage comprising 100 iterations. The $M^2$ and PER values of the selected $\mathbf{x}_0$ for the subsequent local domain are marked as red and gray diamonds, respectively.
        \textbf{c--f}, Experimentally measured near-field (NF) intensity images in horizontal and vertical linear polarizations, and far-field (FF) images in horizontal polarization, corresponding to the reference vectors marked by (c,d,e,f) in (\textbf{b}). Progressive improvement in $M^2$, PER, and the field correlation with the target output $C_{\mathbf{b}}$ is observed.
    }\label{fig:ExpOptimization}
\end{figure}

We perform \textit{in situ} optimization of the MMF amplifier output at \qty{538}{\W} based on the proposed framework. As a representative example, we present the convergence behavior and intermediate beam profiles for a H-pol Gaussian output beam in Fig.~\ref{fig:ExpOptimization}. The overall procedure follows a standard gradient-based optimization scheme: at each iteration, the input vector field $\mathbf{x}$ is updated via the adjoint TM (either the global or local TM), and the corresponding output vector field $\mathbf{y}$ is measured. We maintain the seed power coupled into the MMF amplifier by optimizing the input spatial wavefronts on the basis of high transmission eigenchannels. This ensures that the input and output powers remain constant during the optimization process. Hence, $\mathbf{x}$ and $\mathbf{y}$ can be considered as normalized complex vectors in an $N_\mathrm{fib}$-dimensional space (Fig.~\ref{fig:ExpOptimization}a). Experimental and technical details are described separately in Methods.

The optimization begins with the global TM measurement. The TM is reconstructed from a series of measurements corresponding to the prepared input vectors (see Methods). A key difference from linear cases, however, is that the TM of our nonlinear MMF amplifier depends on the choice of sampling vectors. For instance, sampling over basis vectors, as in conventional linear TM measurements \cite{popoff2010measuring}, may lead to a biased TM that does not fully capture the linear regression of the general input--output mapping. For this reason, we use random vectors for TM measurements (see Methods). This approach also prevents amplifier damage due to self-lasing by maintaining the seed power coupled into the MMF core throughout the measurement process.

Once the global TM is measured, we set the initial solution for the input $\mathbf{x}$ as the linear pseudoinverse of the target output $\mathbf{b}$ (see Methods). If the system were linear (regardless of its complexity), this would directly provide the optimal input solution $\mathbf{x}_*$. Due to strong nonlinearity, however, we observe a significant deviation of the amplifier output from a linearly polarized Gaussian beam, resulting in $M^2=3.95$ and $\mathrm{PER}=\qty{4.6}{\dB}$ (Fig.~\ref{fig:ExpOptimization}c). $M^2$ is calculated from the second moments of the NF and FF intensities of H-pol \cite{siegman1990new}, and $\mathrm{PER}$ from the spatially-integrated NF intensities of H-pol to V-pol.

Next, we perform gradient-based optimization using the global TM. Since the gradient is estimated from the experimentally measured displacement from the target (i.e., the residual $\mathbf{r}=\mathbf{b}-\mathbf{y}$, see Methods), the iterations initially lead to a decrease in $M^2$ and an increase in $\mathrm{PER}$. However, after 20--30 iterations, $M^2$ and $\mathrm{PER}$ begin to stagnate and the optimization does not converge. This behavior indicates that our MMF amplifier is a highly nonlinear system whose gradients cannot be adequately approximated by the global TM. The optimization is stopped after 100 iterations without convergence. The best result we achieve in this stage is shown in Fig.~\ref{fig:ExpOptimization}d with $M^2=2.2$ and $\mathrm{PER}=\qty{9.1}{\dB}$.

Then we switch to the local TM with a reduced input domain size. Since $\mathbf{x}$ is a normalized complex vector in $N_\mathrm{fib}$ dimensions, the global input space lies on a complex unit hypersphere, as shown in Fig.~\ref{fig:ExpOptimization}a. Accordingly, a local domain can be represented as a spherical cap, with the reference vector ($\mathbf{x}_0$) and the domain size parameter ($\rho$) serving as the pole and base radius, respectively. We choose the reference vector $\mathbf{x}_0$ to be the input that provides the best result from the preceding optimization. We use the output field overlap with the target pattern, $C_{\mathbf{b}}\equiv|\mathbf{y}^\dagger \mathbf{b}|\in[0,1]$, as a figure of merit (see Methods). The choice of $\rho$, however, requires more careful consideration: a smaller $\rho$ is preferred for better convergence with local TM, yet it must be large enough such that the solution $\mathbf{x}_*$ remains within the local domain. This requirement leads us to another \textit{in situ} optimization for $\rho$ (see Methods). Since $\mathbf{x}_*$ is unknown in the input space, the validity of $\rho$ should instead be tested in the output space, where $\mathbf{b}$ is given. Specifically, at a given $\rho$, we experimentally determine the boundary of the output region and confirm that $\mathbf{b}$ falls within it (see Methods). This ensures the presence of $\mathbf{x}_*$ in the local domain as long as the nonlinear input--output mapping is a smooth and continuous function (Fig.~\ref{fig:1Overview}b). After the determination of $\rho$, we measure the local TM. The acquisition and reconstruction procedures are mostly identical to those of the global TM, except that the random input vectors are now chosen from the local domain (see Methods).

As shown in Fig.~\ref{fig:ExpOptimization}b, the local TM significantly facilitates convergence of gradient-based optimization, indicating that it captures a more accurate gradient toward the solution than the global TM. However, a single local TM stage is usually not sufficient to reach the solution, and $M^2$ starts increasing and PER starts decreasing near the end of 100 iterations. We therefore select the best result among the 100 iterations and repeat the local domain definition around it, followed by local TM measurement and gradient-based optimization. After one global and six local TM stages, we finally obtain a Gaussian beam with $M^2=\num{1.09}$ and $\mathrm{PER}=\qty{19.6}{\dB}$ as the experimental output of our MMF amplifier at \qty{538}{\W} (Figs.~\ref{fig:ExpOptimization}e--f). The number of local TM stages depends strongly on the beam quality criteria, which are set to $M^2<1.1$ and $\mathrm{PER} > \qty{15}{\dB}$ here. For other structured beam shapes shown below, the output field correlation $C_{\mathbf{b}}>0.95$ is used as a general criterion, which corresponds to more than $90\%$ of the total power contained in the target pattern. Had we applied the criterion of $C_{\mathbf{b}}>0.95$ to the Gaussian beam, the optimization would have ended at the second local TM stage, yielding the result shown in Fig.~\ref{fig:ExpOptimization}e with $M^2=\num{1.17}$ and $\mathrm{PER}=\qty{18.0}{\dB}$.

As indicated at the top of Fig.~\ref{fig:ExpOptimization}b, the local domain size ($\rho$) progressively decreases across successive stages (from 0.53 to 0.17) as expected. However, since $\rho$ is determined from measured data with noise, it may increase during the optimization, especially after an unsuccessful stage. An example is shown in the last stage of Fig.~\ref{fig:ExpOptimization}b, where $\rho$ increases from 0.17 to 0.3 after the preceding local TM stage exhibits a moderate decrease in PER.

A key distinction between our method and numerical gradient estimation methods \cite{nocedal2006numerical, lin2021high} is that the measured TM is an overall approximation of the input--output mapping within the local domain, but the numerical gradient is evaluated at a specific point $\mathbf{x}$. Hence, the local TM is much more robust to a noisy environment than numerical gradient estimators which are highly sensitive not only to experimental random noise but also to the choice of step size \cite{nocedal2006numerical}. Trust-region methods employ a local quadratic model to facilitate convergence \cite{nocedal2006numerical}, but they typically require the nonlinear mapping to be a known function whereas our method does not.

\subsection*{Structured light at 538 W}\label{subsec:Result}
\begin{figure}[t]
\centering
\includegraphics[width=1.0\textwidth]{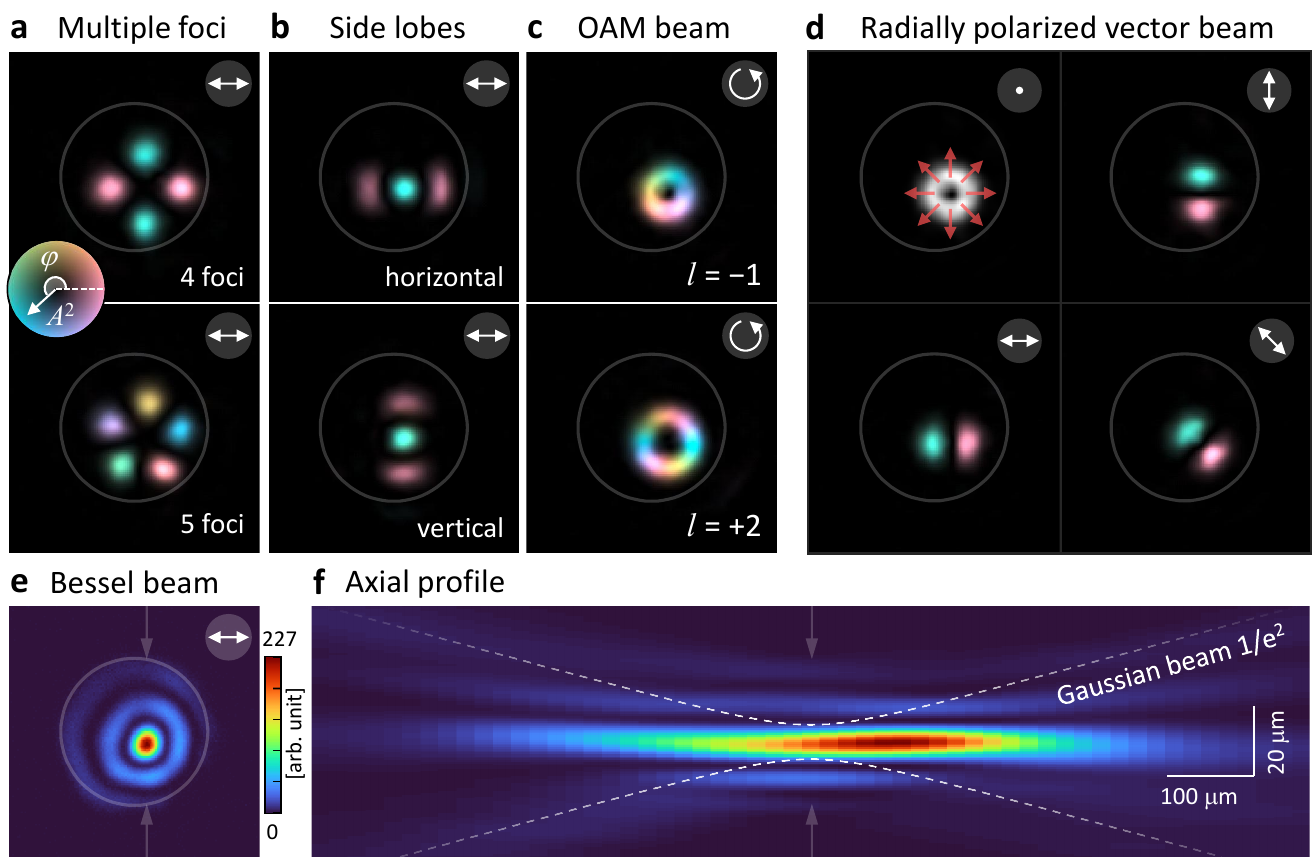}
\caption{
    \textbf{Experimentally generated structured beams at \qty[text-series-to-math]{538}{\W}.}
    \textbf{a--c}, Six different near-field (NF) beam shapes from the MMF amplifier displayed with intensity ($A^2$) and phase ($\varphi$) represented by lightness and chroma, respectively. The overlaid gray circle denotes the fiber core, whose diameter is \qty{42}{\um}. Four- and five-foci beams (\textbf{a}), a bright central spot with two side lobes arranged horizontally and vertically (\textbf{b}), and orbital angular momentum (OAM) beams with topological charges of $l=\numlist[retain-explicit-plus]{-1;+2}$ (\textbf{c}) are created by input wavefront shaping. The OAM beams are generated in left-handed circular polarization, and the others in horizontal linear polarization.
    \textbf{d}, Radially polarized vector beam. The top-left panel shows the polarization-summed NF intensity image, with red arrows indicating the local polarization direction. Three polarization-resolved images show intensity (lightness) and phase (chroma) distribution, with the corresponding polarization direction denoted by white arrows in the top-right corner of each image.
    \textbf{e--f}, Measured NF intensity distribution of a H-pol Bessel beam (\textbf{e}) and its axial intensity profile obtained from numerical propagation of the measured field distribution (\textbf{f}). The gray arrows mark the intersection between transverse and longitudinal profiles in \textbf{e} and \textbf{f}. The white dashed lines in \textbf{f} indicate the $1/e^2$ width contour of a Gaussian beam of the same size as the main lobe of the Bessel beam for comparison. Measured NF and FF images in different polarizations for all beam shapes are presented in Fig.~\ref{figS:rawData}.
}\label{fig:Results}
\end{figure}
The first and perhaps most important beam shape we generate from the high-power MMF amplifier is the Gaussian beam with high beam quality and polarization purity, as shown in Fig.~\ref{fig:ExpOptimization}f. Since detrimental nonlinear effects and instabilities that limit power scaling \cite{Cesar2013high, zervas2014high} are significantly weaker in MMF amplifiers \cite{chen2023mitigating, wisal2024theorySBS, rothe2025wavefront, chen2023suppressing, wisal2024theoryTMI}, they can reach much higher power levels than single-mode and few-mode fiber amplifiers. The main concern for MMF amplifiers has been their poor output beam quality. Our demonstration of an output beam with $M^2<\num{1.1}$ and $\mathrm{PER}\simeq\qty{20}{\dB}$ at \qty{538}{\W} highlights the promise of MMF amplifiers for higher-power single-frequency lasers.
     
In addition to Gaussian beams, the MMF amplifier can produce various types of structured light needed for different applications. Since suppression of nonlinear effects such as SBS and transverse mode instability (TMI) in the MMF amplifier relies on distributing the signal power across a large number of fiber modes \cite{chen2023mitigating, wisal2024theorySBS, wisal2024theoryTMI, rothe2025wavefront}, we ensure that the target patterns to span a sufficient number of modes in the MMF by off-centering them in the NF and/or FF (see Methods, Fig.~\ref{figS:Pnum}).

For parallel material processing and optical sensing, simultaneous generation of multiple focal spots at chosen locations is useful. Using the local TM framework, we shape the seed to the MMF amplifier to generate multiple foci with preset phase values. The output NF images are shown in Fig.~\ref{fig:Results}a, with the phase distribution retrieved from the measured FF images (see Methods, Fig.~\ref{figS:rawData}). Distinct phase values are observed in both four- and five-foci beams, with phase differences between the nearest-neighbor foci of $\pi$ and $0.8\pi$, respectively. Uneven intensity distributions are also demonstrated in Fig.~\ref{fig:Results}b. A brighter central spot and two dimmer side lobes are generated along both horizontal and vertical directions. The side lobes benefit metal welding and powder bed fusion by reducing local overheating and spattering.
    
OAM beams have been of particular interest among high-power structured beams due to their unique ability to deliver optical torque \cite{harrison2024progress, carbajo2025structured}. For example, OAM beams can introduce local chirality in laser material processing. 
Similarly, in laser ablation, OAM beams have been shown to produce smoother and deeper processed surfaces, attributed to the rotational motion of molten material and laser-induced plasma. 
We generate high-power OAM beams with topological charges $l=\numlist[retain-explicit-plus]{-1;+2}$ directly from the MMF amplifier (Fig.~\ref{fig:Results}c). A clear azimuthal phase progression is observed, with its direction and the number of phase windings determined by the sign and absolute value of $l$, respectively. Here, the OAM beams are generated in left-handed circular polarization, demonstrating controllability over polarization states for spin--orbit interactions of light \cite{bliokh2015spin}.  


We further generate vector beams, whose polarization state varies across space. As an example, Fig.~\ref{fig:Results}d shows a radially polarized beam from the MMF amplifier. The polarization-summed image exhibits a doughnut-shaped intensity profile, similar to that of OAM beams (top-left, Fig.~\ref{fig:Results}d). The radial polarization state is confirmed by the linearly polarized images (top-right and bottom, Fig.~\ref{fig:Results}d). A phase difference of $\pi$ between the two lobes is observed, indicating that their polarization directions are opposite. The measured local polarization directions are overlaid on the polarization-summed image. As the radially polarized beam can provide a sharper focus than any scalar beam shape \cite{dorn2003sharper}, it has been utilized for high-precision cutting and microdrilling applications \cite{harrison2024progress}. 

For laser drilling, dicing, and cutting of volumetric materials, an axially extended focus is useful \cite{duocastella2012bessel}. We thus generate a Bessel beam with an extended focal depth from the MMF amplifier. The Bessel beam exhibits multiple rings around the central spot (Fig.~\ref{fig:Results}e), but it is not propagation invariant due to the finite size of the fiber core. Nevertheless, the depth of focus is significantly extended compared with the Gaussian beam, as highlighted in Fig.~\ref{fig:Results}f. The Bessel beam is also useful for one-dimensional optical trapping of cold atoms and atomic waveguides \cite{carbajo2025structured}. 

\section*{Discussion and Conclusion}\label{sec:Conclusion}
Because the local TM is retrieved from experimental data, it provides a robust linear approximation of the high-power MMF amplifier without \textit{a priori} knowledge of the governing equations, system complexity, or practical error models \cite{wright2022nonlinear}. However, the validity of the TM relies on setup stability, which is also important for the \textit{in situ} optimization process. To ensure thermal stability, the amplifier system was operated at the target power for at least 1 hour prior to optimization. To minimize perturbation from the external environment, the entire system is enclosed and isolated, so that we can experimentally reproduce the same output for a given input. Once the amplifier system is stabilized, the input SLM patterns for various output beam shapes can be pre-optimized, stored, and subsequently uploaded on demand. 

Beyond generating static beam profiles, our framework supports real-time reconfiguration of the output beam shape. This capability is particularly valuable for free-space and underwater optical communications \cite{ke2024adaptive} as well as volumetric laser material processing \cite{salter2019adaptive}, where on-the-fly adaptive compensation for environment- or object-induced optical aberrations is important. In the current demonstrations, however, the full optimization of a structured beam takes 15--50 mins, limited largely by the slow response of the liquid-crystal SLM (see Methods). We expect that the introduction of faster beam shapers, such as kHz-rate phase light modulators \cite{rocha2024fast}, will significantly accelerate \textit{in situ} optimization. Furthermore, our scheme may be combined with deep learning by training a neural network with the input--output dataset from optimization, enabling fast inference by incorporating rare target outputs into the training set.

As shown in Fig.~\ref{fig:Setup}b, we employ phase grating patterns on the SLM to modulate the amplitude and phase of both polarizations. Our vector field control scheme is stable and robust, but at the expense of power loss (\qty{-7.4}{\dB}). We emphasize, however, that for a given fractional power loss, the absolute optical power lost in shaping a low-power input is much smaller than that of the high-power output. Compared to structured light lasers, our MMF amplifier shares the same advantage of direct generation of structured light at the source, but our scheme is applicable over a much broader range of power levels. Intra-cavity generation of structured light typically relies on selective mode lasing, which is inherently power sensitive. The MMF amplifier can produce structured light at varying output power levels, with the input wavefront re-optimized at each pumping level.

In conclusion, we demonstrate the direct generation of high-power structured light from a single-frequency MMF amplifier by shaping the low-power seed. The versatility of our system is showcased by the generation of diverse beam shapes with distinct amplitude, phase, and polarization distributions. Our approach enables real-time, on-demand, and adaptive control of the output beam shape without high-power handling. It can therefore be extended to larger-core MMF amplifiers with kilowatt-level output power, with broad implications for high-power laser-based applications. Beyond multimode fiber amplifiers, the local transmission matrix framework provides a general strategy for controlling nonlinear complex systems for which theoretical models are unavailable or inaccurate. We expect it to enable more precise and robust control of various nonlinear optical processes such as stimulated Raman scattering \cite{thompson2016wavefront, tzang2018adaptive}, four-wave mixing \cite{tzang2018adaptive, deliancourt2019wavefront, hary2023tailored}, second-harmonic generation \cite{yao2012controlling, frostig2017focusing}, terahertz generation \cite{cecconi2023nonlinear}, and parametric down-conversion \cite{lib2020real}. By combining experimental sampling with local linearization, our control approach fills the gap between model-based and data-driven control, offering a scalable route to manipulating complex wave systems across optics and beyond.

\clearpage
\section*{Methods}\label{sec:Methods}

\subsection*{Experimental setup}\label{subsec:expSetup}
The detailed experimental setup is shown in Fig.~\ref{figS:Setup}. The continuous-wave seed is provided by a single-frequency linearly polarized fiber laser (RFLM-100-1-1064, NP Photonics Inc.) at $\lambda_s=\qty{1064}{\nm}$ with a \qty{17}{\kHz} linewidth (\qty{-20}{\dB}) \cite{rothe2025wavefront}. The seed is launched into a two-stage preamplifier composed of Yb-doped single-mode fibers and amplified to \qty{45}{\W}. The preamplified beam then passes through a flat-top beam shaper ($\pi$Shaper\_12\_12\_1064\_HP\_W, AdlOptica Optical Systems GmbH) before being directed onto an SLM (SLM-300, Santec Holdings Corp.). The flat-top beam shaper ensures a uniform intensity distribution across the SLM active area without power loss.

The first-order diffraction of the SLM is used to eliminate the contribution of the unmodulated beam (i.e., the zeroth order). As shown in Fig.~\ref{fig:Setup}a, the first order consists of two suborders corresponding to two polarization states. A half-wave plate rotates the polarization of one suborder such that the two suborders have perpendicular polarizations, and they are subsequently combined into a single vector field using a beam displacer (BD27, Thorlabs, Inc.). The combined vector field is conjugated to the back focal plane of a high-power objective lens (LMH-5X-1064, Thorlabs, Inc.) and coupled into the signal fiber, a passive triple-clad multimode fiber (FUD-4693, Coherent Corp.). Its core, first cladding, and second cladding have diameters of \qtylist{42;570;650}{\um}, respectively, with numerical apertures (NAs) of \numlist{0.1;0.22;0.46}. The core supports $N_\mathrm{fib}=80$ modes (40 modes per polarization) at the signal wavelength of \qty{1064}{\nm}. The input facet of the fiber is cleaved at an angle of \qty{8}{\degree} to suppress spurious back-reflections. 

Five fiber-coupled laser diodes each delivering \qty{150}{\W} of optical power at $\lambda_p=\qty{976}{\nm}$ serve as the pump source. They are combined with the signal fiber via a (6+1)$\times$1 side-pump signal-pump combiner (SPC). The pump light is guided by the first cladding, which has a designed absorption coefficient of \qty{3.4}{\dB/\m} at $\lambda_p$. The matching Yb-doped active fiber (FUD-4715, Coherent Corp.) is then spliced to the combiner output. The active fiber is \qty{4.8}{\m} long and mounted on a water-cooled metal plate. The distal end of the Yb-doped fiber is spliced to a short passive fiber with the same parameters as the signal fiber, and terminated with an antireflection-coated end cap (FEC5-1064, Thorlabs, Inc.). To monitor possible self-lasing and SBS-induced instability in the MMF amplifier, backward-propagating light is routed to an optical spectrum analyzer (AQ6370D, Yokogawa Test \& Measurement Corp.) and an amplified photodetector (PDA20CS, Thorlabs, Inc.) connected to an oscilloscope (DSOX3014T, Keysight Technologies, Inc.).

After the end cap, a diffraction-limited aspheric lens (AL5040H-B, Thorlabs, Inc.) collimates the high-power output. A dichroic mirror (DMLP1000L, Thorlabs, Inc.) decouples the residual pump from the signal, and their powers are measured separately by two water-cooled power meters (L250W-BB-50 and L1500W-BB-50, Ophir Optronics Solutions Ltd.). Two beam samplers (BSF20-C, Thorlabs, Inc.) are placed before the signal power meter to direct a small amount of signal to characterize the output beam profile. A half-wave plate is inserted between the two beam samplers to compensate for their polarization-dependent reflectance. A bandpass filter (FLH1064-3, Thorlabs, Inc.) placed after the beam samplers ensures complete rejection of any residual pump light.

A doublet lens (AC254-300-C, Thorlabs, Inc.) serves as the imaging tube lens, providing 7.5$\times$ magnification. The magnified image of the interface between the fiber and the end cap is projected onto a \qty{550}{\um} pinhole, which rejects residual cladding-guided signal and ambient background. The light is then re-collimated and split into two arms to simultaneously observe the near- and far-fields. The near-field arm incorporates an additional imaging tube lens, while the far-field arm directly observes the collimated beam. In both arms, horizontally and vertically polarized images are acquired in a single shot by placing a beam displacer in front of the camera (MQ013RG-ON-S7, XIMEA GmbH). The observed polarization states can be switched to left- and right-handed circular polarizations by introducing a quarter-wave plate using a liquid crystal retarder (LCR) positioned before the beam splitter.

\subsection*{Global TM measurement}\label{subsec:GlobalTM}
We reconstruct the TM of the MMF amplifier directly from near- and far-field intensity images using reference-free phase retrieval algorithms \cite{lee2016exploiting, luo2020phase}. Unlike common-path TM measurement schemes that use a portion of the input as a reference \cite{popoff2010image, yoon2015measuring}, individual complex-valued TM elements are retrieved without a speckled reference beam. The TM measurement procedure consists of four steps: (i) defining the input basis, (ii) generating random input vectors, (iii) measuring the output near- and far-field intensity images, and (iv) reconstructing the TM from the intensity measurements. Note that steps (i) and (ii) do not affect the measurement time in practice, as they can be performed \textit{a priori} on the computer. In this section, we consider global TM measurements only; local TM measurements are discussed separately below.

In step (i), we define a hexagonal lattice at the input fiber facet to serve as the input basis. First, the lattice constant $a$ is determined to satisfy the condition $\lambda_s/\mathrm{NA}_\mathrm{co} > \sqrt{3}a$ (with the fiber core NA  $\mathrm{NA}_\mathrm{co}=0.1$), ensuring that the Brillouin zone of the lattice fully encompasses the numerical aperture of the fiber core. Next, the number of lattice points ($N/2$, where $N$ denotes the total number of points for both polarizations) is set such that the spatial extent of the lattice covers the fiber core. To ensure that all fiber-guided modes are captured, a safety factor of 1.25 is applied to both the fiber NA and the core diameter, yielding a lattice constant of $a=\qty{4.9}{\um}$ and $N/2=151$ lattice points.

Since the beam displacer output facet is conjugated to the fiber input facet, the lattices for the two polarizations generated by the SLM are displaced such that they are merged by the beam displacer. The image aberrations introduced by the beam displacer for each polarization are analytically computed, and the corresponding correction patterns are applied to the SLM. For both polarizations, the manufacturer-provided SLM correction pattern and a linear phase ramp for the first diffraction order are additionally applied. Note that compensating for setup aberrations is not strictly necessary, as the TM will ultimately account for all aberration effects. However, precompensating for aberrations reduces the required safety margin in the lattice design and the total number of measurements needed.

In step (ii), we generate random input vectors based on the defined input basis. Let the basis vectors be $\mathbf{e}_1,\mathbf{e}_2,\ldots,\mathbf{e}_N$, where each $\mathbf{e}_n$ represents a lattice point on the fiber facet. Each random input pattern takes the form $\sum_{n=1}^N x_n\mathbf{e}_n$, where $x_n$ is a complex-valued random coefficient satisfying $\sum_{n=1}^N \left|x_n\right|^2=1$. However, generating such random vectors using a phase-only SLM generally results in a non-uniform amplitude distribution on the SLM, leading to additional power loss. To minimize this loss, we introduce a simple iterative algorithm to find random vectors $\mathbf{x}= \left[ x_1,x_2,\ldots,x_N \right]^\mathrm{T}$ that have a uniform amplitude distribution on the SLM plane. See Algorithm~\ref{algo:phaseOnlyXmat} for details.

In step (iii), we measure the output near- and far-field images corresponding to a series of random input vectors. We first define the complex-valued near-field output as $\mathbf{y}\in\mathbb{C}^{M}$, where $M$ is the number of pixels on the near-field camera. Note that $\mathbf{y}$ spans both polarizations due to the beam displacer placed in front of the camera. The TM captures the linear relation between all input and output fields,
\begin{equation} 
    \mathbf{Y} = \mathbf{T}\mathbf{X},
    \label{eq:TM}
\end{equation}
where $\mathbf{T}\in\mathbb{C}^{M\times N}$ is the TM, $\mathbf{X}=\left[\mathbf{x}_1,\mathbf{x}_2,\ldots,\mathbf{x}_K\right]$ and $\mathbf{Y}=\left[\mathbf{y}_1,\mathbf{y}_2,\ldots,\mathbf{y}_K\right]$ are the input and output matrices, $\mathbf{y}_k$ is the $k$-th output field corresponding to the $k$-th random input $\mathbf{x}_k$, and $K$ is the number of measurements. The measured near- and far-field intensity images are $\mathbf{N}=\left[\mathbf{n}_1,\mathbf{n}_2,\ldots,\mathbf{n}_K\right]$ and $\mathbf{F}=\left[\mathbf{f}_1,\mathbf{f}_2,\ldots,\mathbf{f}_K\right]$, respectively, where $\mathbf{n}_k=\mathbf{y}_k^*\odot\mathbf{y}_k$ is the $k$-th near-field intensity pattern, $\mathbf{f}_k= 
\mathcal{F}\left\{\mathbf{y}_k\right\}^* \odot
\mathcal{F}\left\{\mathbf{y}_k\right\}$ is the $k$-th far-field intensity pattern, $\odot$ denotes the Hadamard (element-wise) product, and $\mathcal{F}\left\{ \cdot \right\}$ is the two-dimensional Fourier transform. We acquire $K=1510$ near- and far-field image pairs, corresponding to an oversampling ratio of $\gamma = K/N = 5$. The oversampling ratio is an important parameter governing phase retrieval fidelity, and it is empirically known that $\gamma \geq 4$ is required for stable reconstruction \cite{lee2016exploiting, luo2020phase}. The acquisition takes approximately \qty{10}{\minute}, primarily constrained by the SLM refresh rate.

In step (iv), we reconstruct $\mathbf{T}$ from the measured $\mathbf{N}$ and $\mathbf{F}$. To facilitate understanding, we transpose both sides of Eq.~\ref{eq:TM} to obtain $\mathbf{Y}^\mathrm{T} = \mathbf{X}^\mathrm{T} \mathbf{T}^\mathrm{T}$, casting $\mathbf{X}^\mathrm{T}$ as a known operator acting on the unknown $\mathbf{T}^\mathrm{T}$. The measurements can then be reformulated as $M$ independent phase-retrieval problems for the complex-valued column vectors of $\mathbf{T}^\mathrm{T}$ with a known sampling matrix $\mathbf{X}^\mathrm{T}$,
\begin{equation} 
    \mathbf{n}'_m = 
    \left( \mathbf{X}^\mathrm{T}\mathbf{t}'_m \right)^* \odot
    \left( \mathbf{X}^\mathrm{T}\mathbf{t}'_m \right)
    , \quad m=1,2,\ldots,M
    \label{eq:phaseRetrievalEquation}
\end{equation}
where $\mathbf{n}'_m$ and $\mathbf{t}'_m$ are the $m$-th column vectors of $\mathbf{N}^\mathrm{T}$ and $\mathbf{T}^\mathrm{T}$, respectively. Note that $\mathbf{n}'_m$ corresponds to the series of near-field intensity values at the $m$-th camera pixel across $K$ input patterns. Although this formulation is consistent with standard phase retrieval, a key advantage here is our ability to design the sampling matrix $\mathbf{X}^\mathrm{T}$. This allows us to construct $\mathbf{X}^\mathrm{T}$ with independent and identically distributed (i.i.d.) random coefficients and apply random-matrix-based phase retrieval algorithms, which have been extensively studied \cite{candes2015phase, wang2017solving,luo2020phase}. This approach is related to that of Ref.~\cite{cheng2023nonconvex}, but incorporates more advanced algorithms for improved robustness and convergence speed. These include an initialization based on the speckle-correlation scattering matrix \cite{lee2016exploiting}, a loss function based on smoothed amplitude flow \cite{luo2020phase}, and an accelerated optimization scheme such as Adaptive moment estimation (Adam) \cite{kingma2014adam}.

Since Eq.~\ref{eq:phaseRetrievalEquation} is invariant under a global phase change of $\mathbf{t}'_m$, the reconstructed $\mathbf{T}$ has arbitrary phase relations between near-field camera pixels. This is a common limitation shared by all common-path TM measurement schemes \cite{popoff2010image, yoon2015measuring}. To determine the global phase relations per polarization, we incorporate the far-field intensity patterns ($\mathbf{F}$) as a regularization term in the phase retrieval algorithm. However, since the far-field information can only constrain the phase relations within each polarization, the cross-polarization phase remains an unknown constant. To complete the TM reconstruction, we retrieve the cross-polarization phase from additional measurements taken in circular polarization states. Additional details can be found in Algorithm~\ref{algo:TMrecon}.

\subsection*{Transmission eigenchannels}\label{subsec:Tchannel}
The linear TM $\mathbf{T}$ provides the transmission eigenchannels of the MMF amplifier without pumping. The eigenvectors of $\mathbf{T}^\dagger \mathbf{T}$ give the input field patterns for individual channels, and the corresponding eigenvalues are their transmittance. For an MMF with $N_\mathrm{fib}$ modes, the maximum number of significant transmission channels is $N_\mathrm{fib}$, with a sharp drop in transmittance for channels of index beyond $N_\mathrm{fib}$ \cite{cao2023controlling}.

\subsection*{SLM pattern optimization}\label{subsec:SLMopt}
Once the global TM $\mathbf{T}$ is retrieved, we redefine the input basis as an orthonormal basis spanning the transmission eigenvectors, $V=\left\{\mathbf{v}_n\right\}$, where $\mathbf{v}_n$ is the $n$-th eigenvector of $\mathbf{T}^\dagger \mathbf{T}$ and $n=1,2,\ldots, N_\mathrm{fib}$. To generate a target input vector field ($\mathbf{x} \in V$ and $\left\| \mathbf{x} \right\|_2=1$) coupled into the fiber core using a phase-only SLM with maximum power efficiency, we introduce an optimization algorithm that finds an SLM pattern minimizing the loss function,
\begin{equation} 
    \mathcal{L} = 
    \left\| 
        \mathbf{z}_V-\left\| \mathbf{z}_V \right\|_2\mathbf{x} 
    \right\|_2^2
    -\alpha_\mathbf{z}\left\| \mathbf{z}_V \right\|_2^2,
    \label{eq:EtoSLMLossFunc}
\end{equation}
where $\mathbf{z}$ is the vector field generated by the phase hologram on the SLM, $\mathcal{P}_V \{\cdot\}$ is the projection operator onto the input basis $V$, $\mathbf{z}_V = \mathcal{P}_V \{ \mathbf{z} \}$ is the effective input vector field that is coupled to the fiber core, and $\alpha_\mathbf{z} = 0.05$ is a regularization parameter. See Algorithm~\ref{algo:SLMopt} for details. The iteration terminates when the normalized field correlation between $\mathbf{z}_V$ and $\mathbf{x}$ exceeds 0.995. Since the SLM pattern must be calculated every time we change the input, we optimized the algorithm to run in approximately \qty{50}{\ms} on a GPU (Quadro RTX 6000, NVIDIA Corp.).

Throughout the paper, however, the optimization time is largely limited by the slow response time of the liquid-crystal-based SLM used, which is around \qty{300}{\ms}. Specifically, the global and local TM measurements take \qty{10}{\minute} and \qty{4}{\minute}, respectively; the local domain size determination takes \qty{0.5}{\minute}; and a hundred gradient-based iterations take \qty{1.5}{\minute}. The full optimization therefore takes \qtyrange{15}{50}{\minute}, depending on the target beam shape and used quality criteria used.

\subsection*{Local TM measurement}\label{subsec:localTM}
For the local TM measurement, the transmission eigenchannels of the measured global TM ($V$, defined above) are used as a scanning basis, instead of the hexagonal lattice. This reduces the number of input basis vectors from $N=302$ to $N_\mathrm{fib}=80$, lowering the number of required measurements and accelerating the optimization process.

The main difference between global and local TM measurement schemes lies in the vector space spanned by the column vectors of $\mathbf{X}$ in Eq.~\ref{eq:TM}. Suppose that we wish to measure the local TM in the vicinity of a normalized reference vector $\mathbf{x}_0 \in V$, as shown in Fig.~\ref{fig:ExpOptimization}a. The local input matrix is defined as $\widehat{\mathbf{X}}=\left[ \widehat{\mathbf{x}}_1,\widehat{\mathbf{x}}_2,\ldots,\widehat{\mathbf{x}}_{K_{\mathrm{loc}}}\right]$ with
\begin{equation} 
    \widehat{\mathbf{x}}_k=\rho\mathbf{x}^\perp_k + \sqrt{1-\rho^2}\mathbf{x}_0,   
    \label{eq:localXmat}
\end{equation}
where $\mathbf{x}^\perp_k$ is the $k$-th normalized random vector orthogonal to $\mathbf{x}_0$, $\rho$ is a domain size parameter, and $K_\mathrm{loc}=\gamma_\mathrm{loc} N_\mathrm{fib}$ is the number of measurements for the local TM. Since the properties of the local TM depend strongly on $\mathbf{x}_0$ and $\rho$, the systematic determination of these two parameters is one of the key steps in our output field optimization algorithm, as discussed in the optimization section below. Note that $\widehat{\mathbf{x}}_k$ traces the boundary of a spherical cap with radius $\rho$ on the complex unit hypersphere, forming the boundary of the local domain of interest (see Fig.~\ref{figS:domainExp}). Sampling along this boundary rather than the entire domain provides greater variation in the output patterns corresponding to different $\widehat{\mathbf{x}}_k$, preventing potential failure in the local TM reconstruction due to small differences in the measured outputs.

The measurement and TM reconstruction procedures follow steps (iii) and (iv) of the global TM reconstruction. However, since the local input matrix $\widehat{\mathbf{X}}$ is no longer uniformly distributed over the entire input space, the TM reconstruction algorithm (Algorithm~\ref{algo:TMrecon}) may be less efficient than in the global case. To prevent potential instability, an increased oversampling ratio of $\gamma_\mathrm{loc}=7$ is used for the local TM, resulting in $K_\mathrm{loc}=\gamma_\mathrm{loc}N_\mathrm{fib}=560$ measurements.

\subsection*{Output field optimization}\label{subsec:dataAcq}
Our output field optimization procedure consists of three major steps: (I) defining the target output, (II) \textit{in situ} gradient-based optimization using the TM, and (III) defining the local domain and measuring the local TM. Steps (II) and (III) are repeated until the output field is optimized. See Algorithm~\ref{algo:outputOpt} for details.

In step (I), we construct the normalized target output field ($\mathbf{b} \in \mathbb{C}^M$, $\left\|\mathbf{b} \right\|_2=1$) based on the fiber modes. We additionally require $\mathbf{b}$ to span a sufficiently large number of fiber modes to avoid nonlinear effects such as stimulated Brillouin scattering \cite{rothe2025wavefront}. Since core-centered shapes generally span only a few modes, we introduce lateral offsets in both the near- and far-fields to increase mode participation (see Fig.~\ref{figS:Pnum}).

In step (II), we iteratively update the input field ($\mathbf{x} \in V$) based on experimental measurements. The initial estimate is calculated via the pseudoinverse of the global TM ($\mathbf{x}=\mathbf{T}^+\mathbf{b}$). In each iteration, $\mathbf{x}$ is displayed on the SLM, and the corresponding near- and far-field images are acquired in linear (horizontal and vertical) and circular (left- and right-handed circular) polarization states by toggling the $\pi/2$ retardance of the liquid crystal retarder (Fig.~\ref{fig:Setup}a). The normalized output field ($\mathbf{y} \in \mathbb{C}^M$, $\left\|\mathbf{y} \right\|_2=1$) is estimated from the measurements using a modified Gerchberg--Saxton (GS) algorithm (Algorithm~\ref{algo:GSalgo}). To mitigate ambiguities in output fields that have identical NF and FF images (e.g., OAM beams with $l=\pm1$), the target $\mathbf{b}$ is used as the initial guess of the modified GS algorithm. The optimization is stopped when the output beam metrics satisfy the stopping criteria. We use the output field correlation $|\mathbf{y}^\dagger \mathbf{b}|>0.95$ as the default criterion, except for the Gaussian beam (Fig.~\ref{fig:ExpOptimization}f), for which the beam propagation factor ($M^2<1.1$) and polarization extinction ratio ($\mathrm{PER}>\qty{15}{\dB}$) are used instead. If the criteria are not met, $\mathbf{x}$ is updated using the gradient computed from the residual $\mathbf{r}=\mathbf{b}-\mathbf{y}$ and $\mathbf{T}^\dagger$. After 100 iterations without convergence, the procedure advances to step (III).

In step (III), we define the reference vector $\mathbf{x}_0$ as the input field that achieves the highest correlation during the last iteration of step (II) (Fig.~\ref{fig:ExpOptimization}b). We then calculate the field correlation $|\mathbf{b}^\dagger\mathbf{y}_0|$ to estimate how close $\mathbf{x}_0$ is to the solution $\mathbf{x}_\ast$, where $\mathbf{y}_0$ is the normalized output corresponding to $\mathbf{x}_0$. Since we require $\mathbf{x}_\ast$ to lie within the local domain, we define the domain boundary such that $\mathbf{b}$ is closer to $\mathbf{y}_0$ than the boundary output vectors $\widehat{\mathbf{y}}$, i.e., $|\widehat{\mathbf{y}}^\dagger \mathbf{y}_0| < |\mathbf{b}^\dagger \mathbf{y}_0|$, where $\widehat{\mathbf{y}}$ is the output field corresponding to a boundary input vector $\widehat{\mathbf{x}}$ defined in Eq.~\ref{eq:localXmat} (see Fig.~\ref{figS:domainExp} for illustration). At the same time, we prefer $\rho$ to be as small as possible to maximize linearity within the domain, as verified in the simulations discussed below. This leads to a one-dimensional optimization over $\rho$,
\begin{equation} 
    \underset{\rho}{\operatorname{minimize}}\ 
    \left|      
    |\widehat{\mathbf{y}}^\dagger(\rho)\mathbf{y}_0|
    -\varepsilon |\mathbf{b}^\dagger\mathbf{y}_0|
    \right|,
    \label{eq:optLocalSize}
\end{equation}
where $\varepsilon=0.95$ is introduced to ensure $|\widehat{\mathbf{y}}^\dagger(\rho)\mathbf{y}_0|<|\mathbf{b}^\dagger\mathbf{y}_0|$. This optimization is solved using the \texttt{fminbnd} function in MATLAB, which implements Brent's method. Once $\mathbf{x}_0$ and $\rho$ are determined, the local TM is measured as described in the previous section, and $\mathbf{T}$ is updated to the local TM to provide a more accurate gradient in step (II).

\subsection*{Numerical simulation of local TM}\label{subsec:simulation}

Numerical simulations of our MMF amplifier are performed to validate the local TM
framework and to quantify the accuracy of the linear TM approximation of the nonlinear
MMF amplifier as a function of the local domain size $\rho$. The simulation follows our
model of MMF amplifiers with gain saturation and pump depletion in Ref.~\cite{rothe2025wavefront}. We construct a set of $K_\mathrm{train}=7N_\mathrm{fib}$ random input vectors within the
local domain boundary defined in Eq.~\ref{eq:localXmat}, 
$\widehat{\mathbf{X}}=\left[ \widehat{\mathbf{x}}_1,\widehat{\mathbf{x}}_2,\ldots,
\widehat{\mathbf{x}}_{K_{\mathrm{train}}}\right]$. For each input vector, we calculate the corresponding MMF amplifier output fields, 
$\widehat{\mathbf{Y}}=\left[ \widehat{\mathbf{y}}_1,\widehat{\mathbf{y}}_2,\ldots,
\widehat{\mathbf{y}}_{K_{\mathrm{train}}}\right]$. The local TM is fitted via linear regression of the input--output mapping obtained numerically: 
$\mathbf{T}_\mathrm{loc} = \widehat{\mathbf{Y}}\widehat{\mathbf{X}}^+$, where $+$ denotes the Moore--Penrose pseudoinverse. The fidelity of the fitted TM is then evaluated using an independent 
set of $K_\mathrm{test}=7 N_\mathrm{fib}$ random vectors ($\bar{\mathbf{x}}_1, \bar{\mathbf{x}}_2,\ldots,\bar{\mathbf{x}}_{K_{\mathrm{test}}}$) within the same local domain. The accuracy of the local TM is quantified by the test error
\begin{equation}
    \epsilon_\mathrm{test} = 
\frac{1}{K_\mathrm{test}}
\sum_{k=1}^{K_\mathrm{test}}
\frac{
\lVert \bar{\mathbf{y}}_k - \bar{\mathbf{y}}_k' \rVert^2
}{
\lVert \bar{\mathbf{y}}_k \rVert 
\lVert \bar{\mathbf{y}}_k' \rVert
},
    \label{eq:testError}
\end{equation}
where $\bar{\mathbf{y}}_k$ is the MMF amplifier output field from numerical simulation and $\bar{\mathbf{y}}_k' = 
\mathbf{T}_\mathrm{loc}\bar{\mathbf{x}}_k$ is the local-TM prediction for the $k$-th test input from the local domain ($\bar{\mathbf{x}}_k$). The calculated $\epsilon_\mathrm{test}$ is shown by blue crosses in Fig.~\ref{figS:simulation} as a function of $\rho$, swept from zero (corresponding to the reference input $\mathbf{x}_0$) to unity (spanning the full input space). The red solid curve represents a fourth-order polynomial fit of the test error.

\subsection*{Data availability}
All data and algorithms needed to reproduce the results and conclusions of the study are present in the main text, Methods, or Supplementary Information. The raw data for all figures are archived in \url{https://doi.org/10.5281/zenodo.19673521}.

\subsection*{Acknowledgments}
We thank A. Douglas Stone, Kabish Wisal, Filipe M. Ferreira, Nathan Vigne, and SeungYun Han for fruitful discussions. We acknowledge the technical support from Daniel Creeden from Sci Lase LLC. This work is supported by the Air Force Office of Scientific Research (AFOSR) under Grant FA9550-24-1-0129.

\subsection*{Conflict of interest}
The authors declare no conflicts of interest.

\subsection*{Contributions}
K.L. developed optimization algorithms and numerical codes. K.L., B.H., and P.A. built the setup and performed experiments. B.H. performed numerical simulations in collaboration with K.L. M.E., S.R., and C.-W.C. contributed to the early stage of this project. H.C. initiated the project and supervised the research. K.L., B.H., and H.C. wrote the manuscript with input from all authors.

\bibliography{bib_MMFamplifier} 

\newpage
\renewcommand{\thefigure}{S\arabic{figure}}
\renewcommand{\thetable}{S\arabic{table}}
\renewcommand{\theequation}{S\arabic{equation}}
\renewcommand{\thealgorithm}{S\arabic{algorithm}}
\renewcommand{\thepage}{S\arabic{page}}

\renewcommand{\theHfigure}{S\arabic{figure}}
\renewcommand{\theHtable}{S\arabic{table}}
\renewcommand{\theHequation}{S\arabic{equation}}
\renewcommand{\theHalgorithm}{S\arabic{algorithm}}

\setcounter{figure}{0}
\setcounter{table}{0}
\setcounter{equation}{0}
\setcounter{algorithm}{0}
\setcounter{page}{1} 
\pagenumbering{gobble}%

\begin{center}
    \section*{Supplementary Information for:\\ \paperTitle}
    
    KyeoReh~Lee,
    Baichuan~Huang,
    Peyman~Ahmadi,
    Mert~Ercan,
    Stefan~Rothe,
    Chun-Wei~Chen
    and Hui~Cao$^\ast$\\
    
    \small$^\ast$Corresponding authors. Email: hui.cao@yale.edu\\
\end{center}

\subsubsection*{This PDF file includes:}
Figs. S1 to S6\\
Algorithms S1 to S6\\


\clearpage
\subsection*{Supplementary Figures}
\begin{figure}[h]
\centering
\includegraphics[width=0.6\textwidth]{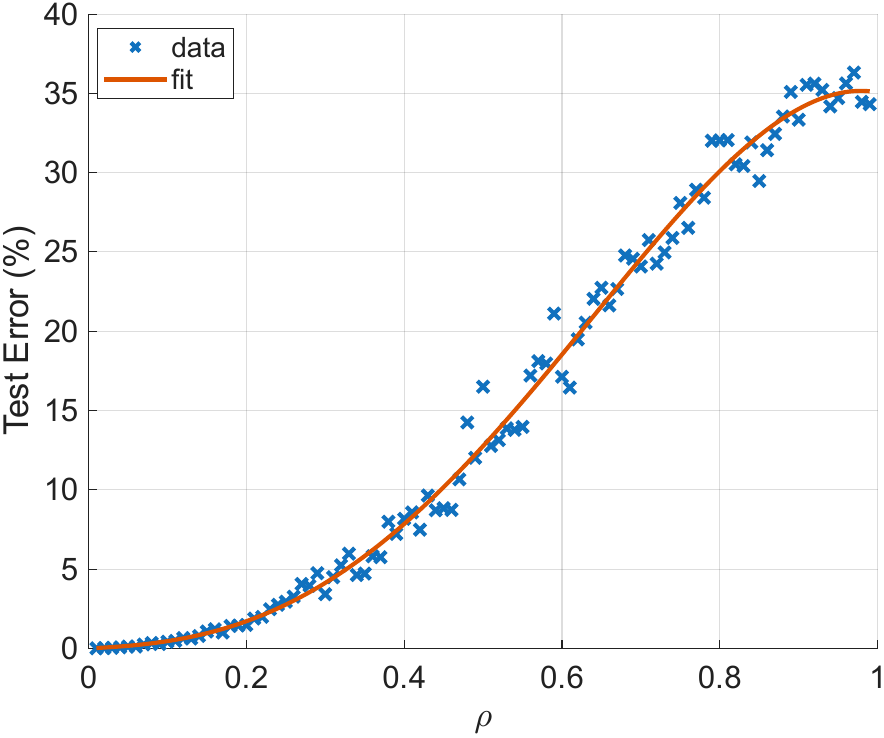}
\caption{
\textbf{Numerical validation of local TM framework.} Test error of a 
linear TM fitted within a local domain of size parameter $\rho$ for our MMF amplifier at \qty{538}{\W} output power. The local domain is defined in Eq.~\ref{eq:localXmat}, where $\rho=0$ corresponds to the reference input $\mathbf{x}_0$ (center of the local domain) and $\rho=1$ spans the entire input space. The test error is defined as the average of $\lVert \bar{\mathbf{y}} - \bar{\mathbf{y}}' \rVert^2 / (\lVert \bar{\mathbf{y}} \rVert \lVert \bar{\mathbf{y}}' \rVert)$, where $\bar{\mathbf{y}}$ and $\bar{\mathbf{y}}'$ are the numerically calculated and local-TM-predicted output fields, respectively, averaged over test inputs that are independent of the inputs used for the local TM regression. The error (blue crosses) grows monotonically with $\rho$, confirming that the 
linear TM approximation becomes increasingly accurate as the local domain contracts. The red solid line is a fourth-order polynomial fit of the numerical test error.
}\label{figS:simulation}
\end{figure}

\begin{figure}[ht]
    \centering
    \includegraphics[width=1.0\textwidth]{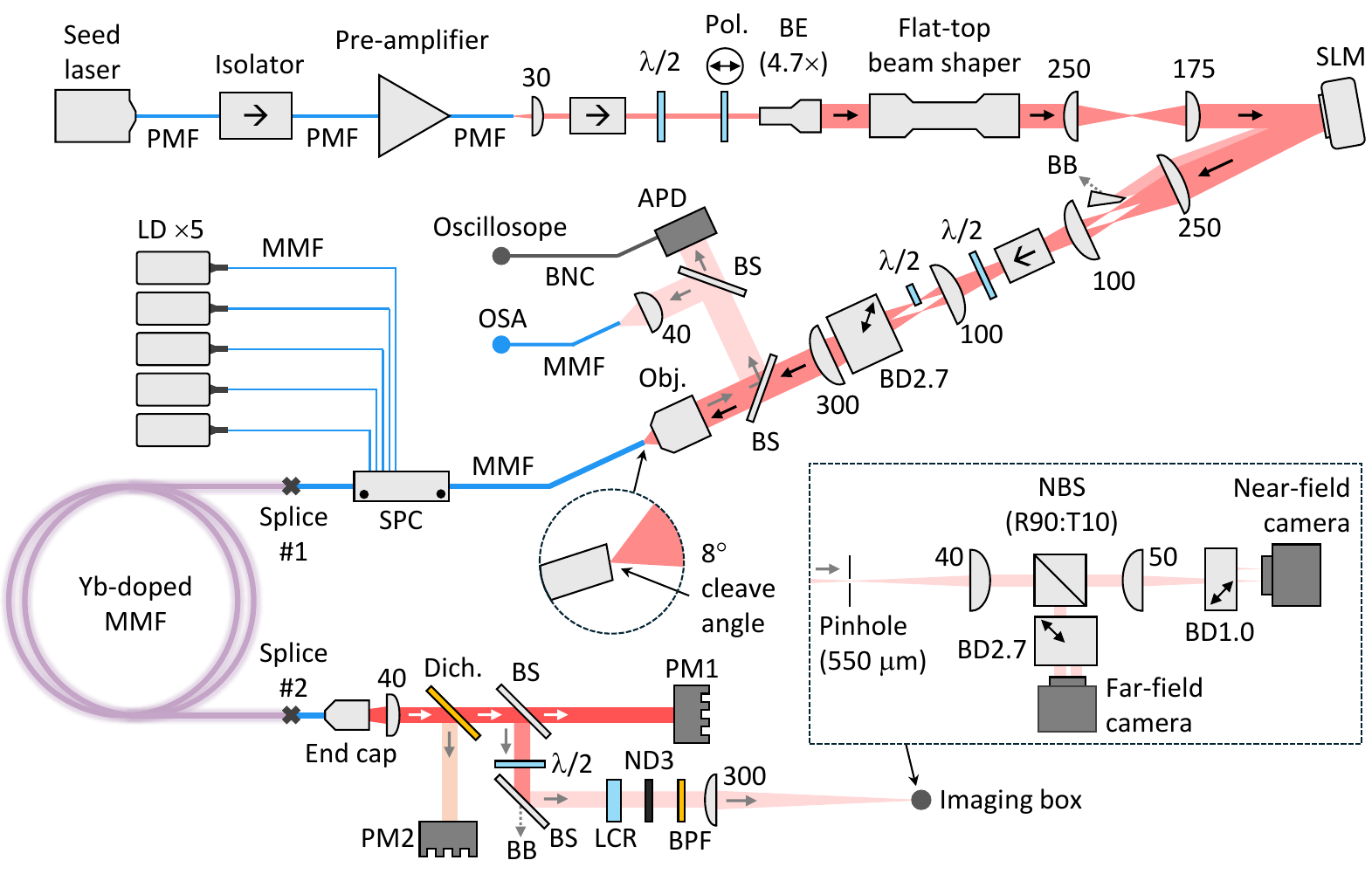}
    \caption{
        \textbf{Detailed experimental setup.}
        Schematic of the full experimental setup. PMF, polarization-maintaining fiber; $\lambda/2$, half-wave plate; Pol., polarizer; BE, beam expander; SLM, spatial light modulator; BB, beam block; BD, beam displacer (the number indicates the beam displacement in mm); Obj., objective lens; MMF, multimode fiber; SPC, signal-pump combiner; LD, laser diode; BS, Fresnel beam sampler; PM, power meter; Dich., dichroic mirror; ND, neutral density filter; LCR, liquid crystal retarder; BPF, bandpass filter; NBS, non-polarizing beam splitter; APD, amplified photodetector; OSA, optical spectrum analyzer; BNC, BNC cable. Numbers next to the lenses denote focal lengths in mm. The inset shows the imaging box containing the near- and far-field cameras. Fiber splice points are marked by crosses.
    }
    \label{figS:Setup}
\end{figure}

\begin{figure}[h]
\centering
\includegraphics[width=0.9\textwidth]{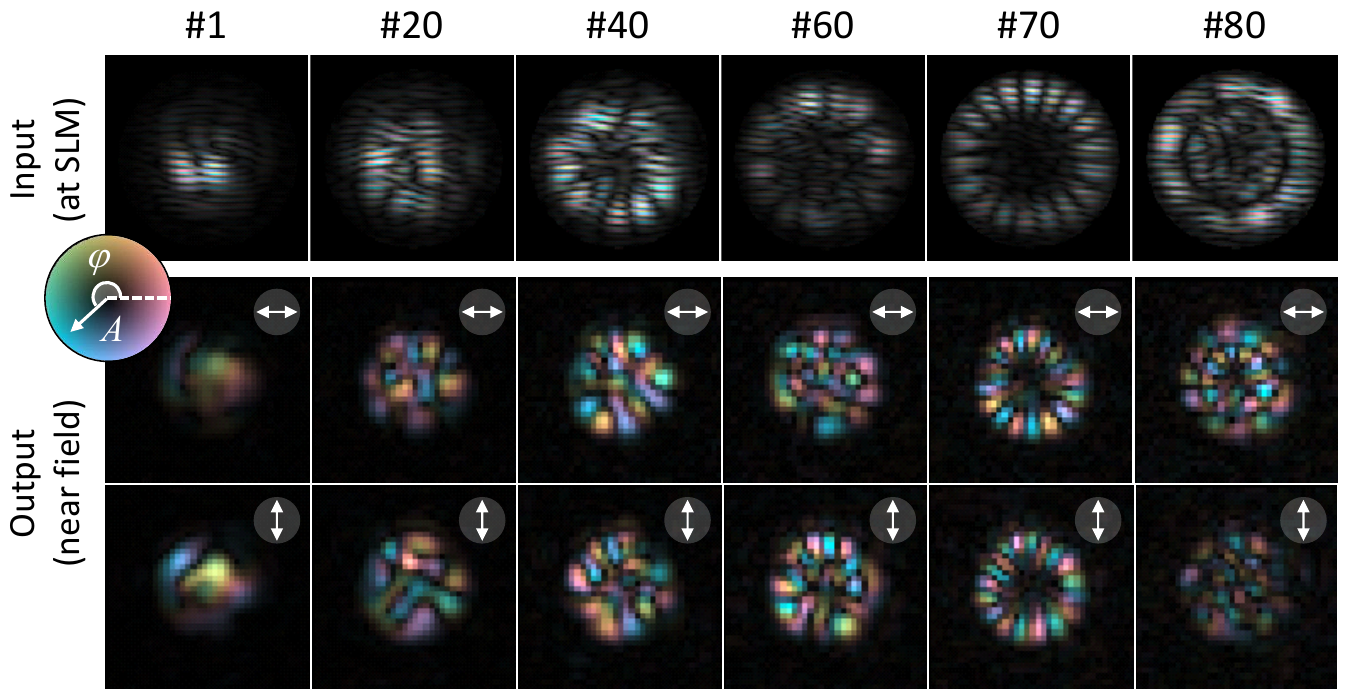}
\caption{
    \textbf{Transmission eigenchannels of MMF without pumping.}
    Selected transmission eigenchannels of the linear MMF amplifier (without pumping), displayed with input and output amplitude ($A$) and phase ($\varphi$) represented by lightness and chroma, respectively. The top row shows the input pattern at the SLM plane. Note that all input patterns exhibit the vertical grating feature seen in Fig.~\ref{fig:Setup}b, which generates two independent suborders corresponding to different polarization states. The bottom two rows show the corresponding output patterns in near-field images in horizontal and vertical linear polarizations (marked by white arrows). Transmission eigenvectors \#1 through \#80 span the $N_\mathrm{fib}=80$ modes in the MMF core and serve as the input basis for the local TM measurement.
}\label{figS:SingularVectors}
\end{figure}

\begin{figure}[h]
    \centering
    \includegraphics[width=1.0\textwidth]{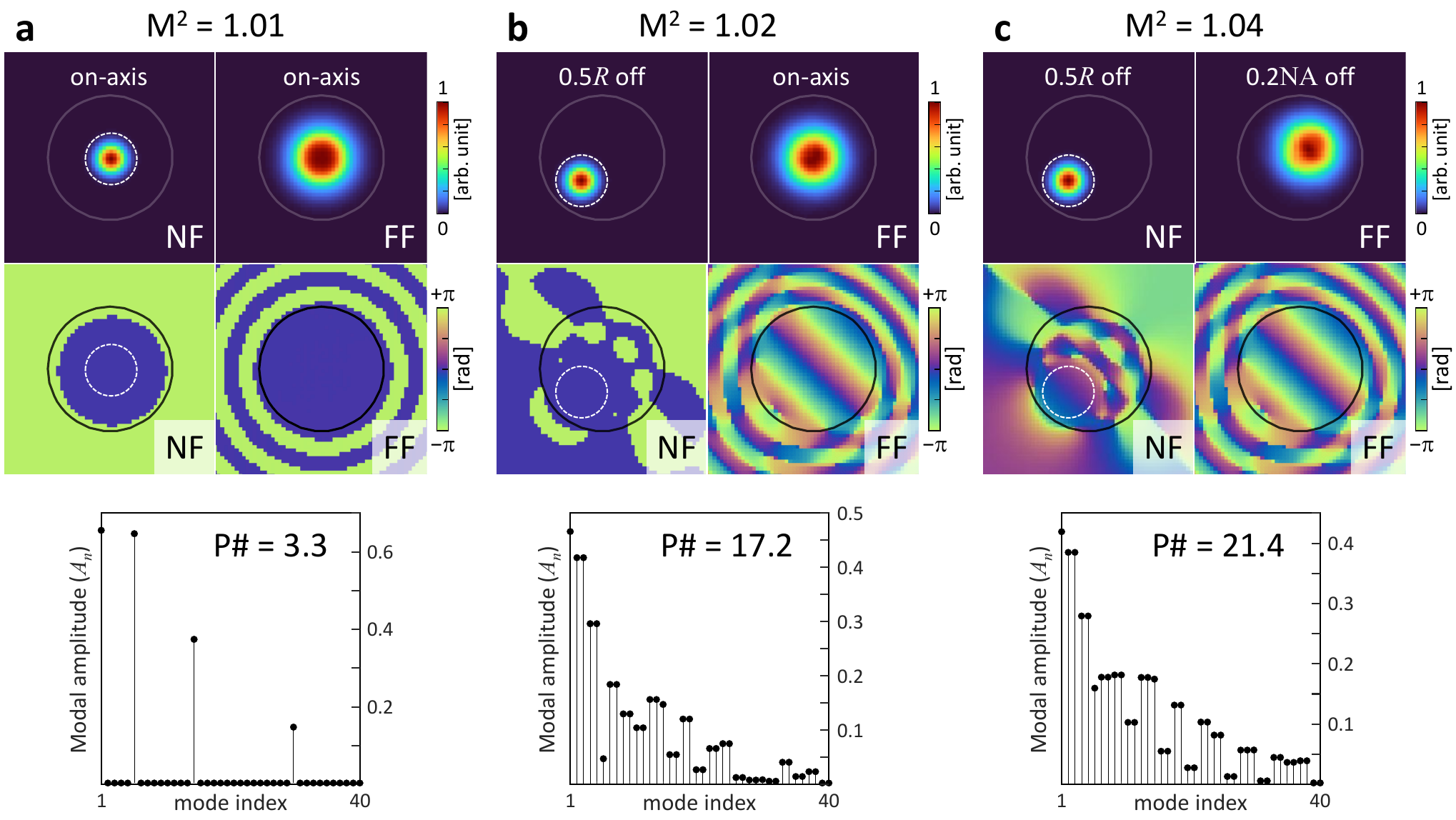}
    \caption{
        \textbf{Off-centering the target pattern to increase mode participation.}
        Simulated Gaussian beam from an MMF with different lateral offsets from the fiber axis in the near-field (NF) and far-field (FF). Each column shows, from top to bottom, the NF and FF intensity distributions, the corresponding phase distributions, and the modal amplitude decomposition ($A_n$) with the participation number, $\mathrm{P\#} = (\sum_n A_n)^2 / \sum_n A_n^2$. $R$ and NA are the core radius and the numerical aperture of the fiber, respectively. The solid black circles in the NF and FF images denote the fiber core and the fiber NA, respectively. The dashed white circles in the NF images denote the Gaussian beam position, where the phase values are meaningful.
        \textbf{a}, On-axis Gaussian beam in both NF and FF ($M^2=1.01$, $\mathrm{P\#}=3.3$). The beam spans only a few low-order fiber modes.
        \textbf{b}, Gaussian beam offset by $0.5R$ from the fiber axis in the NF while remaining on-axis in the FF. The mode participation increases significantly ($\mathrm{P\#}=17.2$) without notable degradation in beam quality ($M^2=1.02$).
        \textbf{c}, Gaussian beam offset by $0.5R$ in the NF and $0.2\mathrm{NA}$ in the FF. The mode participation is further increased ($\mathrm{P\#}=21.4$) and $M^2=1.04$.
    }\label{figS:Pnum}
\end{figure}

\begin{figure}[h]
    \centering
    \includegraphics[width=1.0\textwidth]{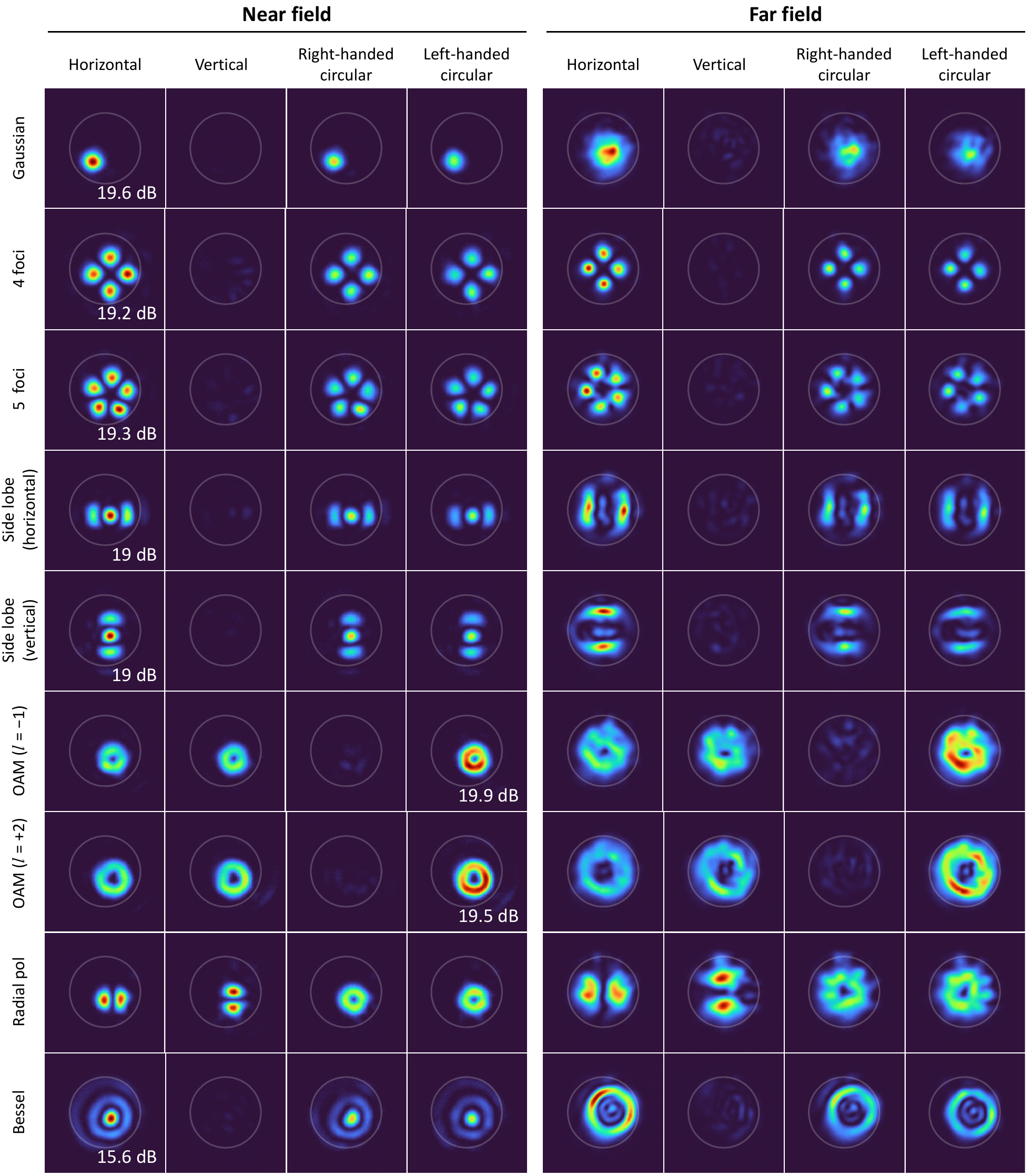}
    \caption{
        \textbf{Raw measurement data for structured beams from the MMF amplifier at \qty{538}{\W}.}
        Near-field (left) and far-field (right) intensity distributions measured at four polarization states (horizontal, vertical, right-handed circular, and left-handed circular) for each optimized beam shape. Rows correspond to, from top to bottom, the H-pol Gaussian beam (Fig.~\ref{fig:ExpOptimization}f), four-foci beam (top, Fig.~\ref{fig:Results}a), five-foci beam (bottom, Fig.~\ref{fig:Results}a), a central bright spot with two horizontal side lobes (top, Fig.~\ref{fig:Results}b) or vertical side lobes (bottom, Fig.~\ref{fig:Results}b), OAM beam in left-handed circular polarization with $l=-1$ (top, Fig.~\ref{fig:Results}c) or $l=+2$ (bottom, Fig.~\ref{fig:Results}c), radially polarized vector beam (Fig.~\ref{fig:Results}d), and H-pol Bessel beam (Fig.~\ref{fig:Results}e). The PER values are indicated for each beam shape at the target polarization state.
    }\label{figS:rawData}
\end{figure}

\begin{figure}[h]
    \centering
    \includegraphics[width=0.6\textwidth]{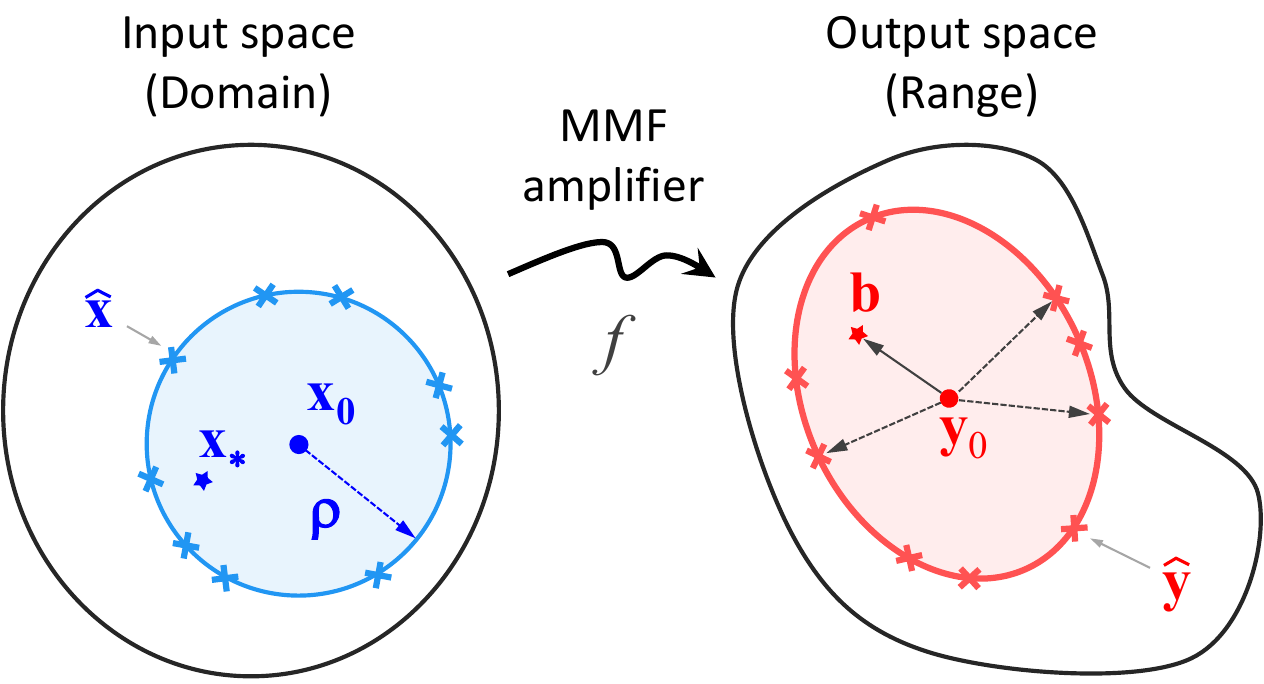}
    \caption{
        \textbf{Selection of the local domain.}
        The global input space (domain) and output space (range) of the MMF amplifier are shown. The local domains are shown as shaded areas in both input and output spaces. In the input space, $\mathbf{x}_*$ is the solution, $\mathbf{x}_0$ is the reference vector, and $\widehat{\mathbf{x}}$ is an input vector on the local domain boundary. Defining the nonlinear input--output mapping of our MMF amplifier as $f:\mathbf{x}\mapsto\mathbf{y}$, the corresponding outputs are $\mathbf{b}=f(\mathbf{x}_*)$ (target), $\mathbf{y}_0 = f(\mathbf{x}_0)$ (output of the reference vector), and $\widehat{\mathbf{y}} = f(\widehat{\mathbf{x}})$ (output of the boundary input vector).
        The local domain size $\rho$ is chosen such that the distance between $\mathbf{b}$ and $\mathbf{y}_0$ (solid arrow) is smaller than the distances between $\widehat{\mathbf{y}}$ and $\mathbf{y}_0$ (dotted arrows), ensuring that $\mathbf{x}_*$ lies within the local domain.
    }
    \label{figS:domainExp}
\end{figure}

\clearpage
\subsection*{Supplementary Algorithms}
\begin{algorithm} [ht]
    \caption{ Phase-only random vector generation. Note that
        $\operatorname{diag}\{ \mathbf{A} \}\in\mathbb{C}^{m \times n}$ is a diagonal matrix made from the main diagonal of $\mathbf{A}\in\mathbb{C}^{m \times n}.$
    } \label{algo:phaseOnlyXmat}
    \begin{algorithmic}[1]
        \State \textbf{Input:} 
        A basis transform matrix $\mathbf{E}=\left[\mathbf{e}_1,\mathbf{e}_2,\ldots,\mathbf{e}_N\right]$, where $\mathbf{e}_n$ is the $n$-th input basis vector; and the oversampling ratio $\gamma$.
        \Statex
        
        \Procedure {PhaseOnlyRandomInputGeneration}{$\mathbf{E},\gamma$} 
        \State $K \gets \gamma N$    
        \For{$k = 1$ to $K$}     
        \State Generate i.i.d. $\mathbf{x}_k \in \mathbb{C}^N\sim
        \mathcal{N}\left(0,\frac{1}{2N}\right)+i\mathcal{N}\left(0,\frac{1}{2N}\right)$
        
        \While{$\mathbf{x}_k$ not converged} 
        \State $\mathbf{z}_k\gets$
        \Call{MakePhaseOnly}{$\mathbf{E}\mathbf{x}_k$}            
        \State $\mathbf{x}_k\gets \mathbf{E}^\dagger \mathbf{z}_k$
        \Comment{ Projection to the input basis }            
        \State $\mathbf{x}_k \gets \mathbf{x}_k/\|\mathbf{x}_k\|_2$
        
        \EndWhile
        \EndFor    
        \State $\mathbf{X} \gets \left[\mathbf{x}_1,\mathbf{x}_2,\ldots,\mathbf{x}_K\right]$    
        \State \textbf{return} $\mathbf{X}$
        \EndProcedure
        \Statex
        
        \Function {MakePhaseOnly}{$\mathbf{x}$} \label{line:MakePhaseOnly}
        \State $\mathbf{x}\gets \operatorname{diag}\{ \mathbf{x}\mathbf{x}^\dagger \}^{-\frac{1}{2}} \mathbf{x}$      
        \State \Return $\mathbf{x}$
        \EndFunction
        
    \end{algorithmic}
\end{algorithm}

\begin{algorithm} [t]
    \caption{ Reference-free TM reconstruction. Please refer to Algorithm~\ref{algo:usedFunc} for detailed steps in the functions used in the iteration. Note that 
        $\operatorname{vec}\{ \mathbf{A}\}\in\mathbb{C}^{mn}$ is the vectorization of a matrix $\mathbf{A}\in\mathbb{C}^{m\times n}$.
    } \label{algo:TMrecon}
    \begin{algorithmic}[1]
        \State \textbf{Input:}
        An input matrix $\mathbf{X}=\left[\mathbf{x}_1,\mathbf{x}_2,\ldots,\mathbf{x}_K\right]$, where $\mathbf{x}_k \in \mathbb{C}^N$; the corresponding near-field images $\mathbf{N}=\left[\mathbf{n}_1,\mathbf{n}_2,\ldots,\mathbf{n}_K\right]$, where $\mathbf{n}_k \in \mathbb{R}_+^M$; the corresponding far-field images $\mathbf{F}=\left[\mathbf{f}_1,\mathbf{f}_2,\ldots,\mathbf{f}_K\right]$, where $\mathbf{f}_k \in \mathbb{R}_+^M$;
        the first 10 columns of $\mathbf{X}$ used for the cross-polarization phase measurement, $\mathbf{X}^{\scriptscriptstyle\lambda/4}=\left[\mathbf{x}_1,\mathbf{x}_2,\ldots,\mathbf{x}_{10}\right]$; the corresponding near-field images measured in circular polarization states, $\mathbf{N}^{\scriptscriptstyle\lambda/4}=\left[\mathbf{n}^{\scriptscriptstyle\lambda/4}_1,\mathbf{n}^{\scriptscriptstyle\lambda/4}_2,\ldots,\mathbf{n}^{\scriptscriptstyle\lambda/4}_{10}\right]$, where $\mathbf{n}^{\scriptscriptstyle\lambda/4}_k \in \mathbb{R}_+^M$; a matrix $\mathbf{P}_{o,e}\in\left\{0,1\right\}^{\frac{M}{2}\times M}$ that extracts the ordinary or extraordinary polarization component of the beam displacer from the near- or far-field images; the step size $\eta = 0.1$; and the regularization parameter $\alpha_\mathbf{f}=0.5$.
        \Statex
        \Procedure {RefFreeTMreconstruction}{$\mathbf{X}, \mathbf{N}, \mathbf{F},
        \mathbf{X}^{\scriptscriptstyle\lambda/4},
        \mathbf{N}^{\scriptscriptstyle\lambda/4},
        \mathbf{P}_o,\mathbf{P}_e,\eta,\alpha_\mathbf{f}$} 
        \State $\mathbf{T}\gets$ 
        \Call{SSMInitialization}{$\mathbf{X}, \mathbf{N}, \mathbf{F}$}
        
        \State $\mathbf{M}_1,\mathbf{M}_2 \gets 0$
        \State $k,\beta_1,\beta_2,\epsilon \gets 0,\,0.9,\,0.999,\,10^{-8}$    
        
        \While{$\mathbf{T}$ not converged}   
        \State $k \gets k + 1$        
        \State $\mathbf{Y} \gets \mathbf{T}\mathbf{X}$
        \State $\mathbf{R} \gets$ 
        $\frac{1}{1+\alpha_\mathbf{f}}$
        \Call{NearFieldResidual}{$\mathbf{Y}, \mathbf{N}$}
        $+\frac{\alpha_\mathbf{f}}{1+\alpha_\mathbf{f}}$ 
        \Call{FarFieldResidual}{$\mathbf{Y}, \mathbf{F}$}
        
        \State $\mathbf{G} \gets -\mathbf{R}\mathbf{X}^\dagger$ 
        \State $\mathbf{M}_1 \gets \beta_1\mathbf{M}_1
        +\left(1-\beta_1\right)\mathbf{G}$
        \State $\mathbf{M}_2 \gets \beta_2\mathbf{M}_2
        +\left(1-\beta_2\right) \left(\mathbf{G}^\ast \odot \mathbf{G}\right)$
        \State $\mathbf{\hat{M}}_1 \gets \mathbf{M}_1/\left(1-\beta_1^k \right)$
        \State $\mathbf{\hat{M}}_2 \gets \mathbf{M}_2/\left(1-\beta_2^k \right)$            
        \State $\mathbf{T}\gets\mathbf{T}-\eta \mathbf{\hat{M}}_1/\left( \sqrt{\mathbf{\hat{M}}_2}+\epsilon \right)$
        \Comment{Adaptive moment estimation (Adam) \cite{kingma2014adam}}
        \EndWhile   
        \State $\varphi \gets$ \Call{CrossPolPhase}{$\mathbf{X}^{\scriptscriptstyle\lambda/4}, \mathbf{T}, \mathbf{N}^{\scriptscriptstyle\lambda/4},\mathbf{P}_o,\mathbf{P}_e$}    
        \State $\mathbf{T}\gets
        \left(
        \mathbf{P}_o^\mathrm{T}\mathbf{P}_o    +\mathbf{P}_e^\mathrm{T}\mathbf{P}_e e^{i\varphi}
        \right)
        \mathbf{T}$
        \Comment{$\mathbf{P}_o^\mathrm{T}\mathbf{P}_o+\mathbf{P}_e^\mathrm{T}\mathbf{P}_e=\mathbf{I}_M$}
        
        \State \textbf{return} $\mathbf{T}$
        \EndProcedure
        \Statex
        \Function {CrossPolPhase}{$\mathbf{X}^{\scriptscriptstyle\lambda/4}, \mathbf{T}, \mathbf{N}^{\scriptscriptstyle\lambda/4}, \mathbf{P}_o,\mathbf{P}_e$}
        \State $\mathbf{y}_{\mathrm{H},\mathrm{V}} \gets \operatorname{vec}\{ \mathbf{P}_{o,e}\mathbf{T}\mathbf{X}^{\scriptscriptstyle\lambda/4}\}$      
        \Comment{$\mathbf{y}_{\mathrm{H},\mathrm{V}}\in\mathbb{C}^{5M}$}
        
        \State $\mathbf{n}_{\mathrm{R},\mathrm{L}} \gets \operatorname{vec}\{
        \mathbf{P}_{o,e} \mathbf{N}^{\scriptscriptstyle\lambda/4} \}$
        \Comment{$\mathbf{n}_{\mathrm{R},\mathrm{L}}\in\mathbb{R}^{5M}_+$}

        \State $\mathbf{a}_0\gets\mathbf{y}_\mathrm{H}^*\odot\mathbf{y}_\mathrm{H} + \mathbf{y}_\mathrm{V}^*\odot\mathbf{y}_\mathrm{V}$
        \State $\mathbf{a}_1\gets i\;\mathbf{y}_\mathrm{H}^*\odot\mathbf{y}_\mathrm{V}$
        
        \State $\mathbf{\Phi}\gets
        \frac{1}{2}\left[
        \mathbf{a}_0,\ +\mathbf{a}_1,\ +\mathbf{a}_1^*;\ 
        \mathbf{a}_0,\ -\mathbf{a}_1,\ -\mathbf{a}_1^*
        \right]$ 
        \Comment{$\mathbf{\Phi}\in\mathbb{C}^{10M \times 3}$}
        
        \State $\mathbf{b}\gets \mathbf{\Phi}^+
        \left[\mathbf{n}_\mathrm{R};\ \mathbf{n}_\mathrm{L}\right]$
        \Comment{$\mathbf{b}=\left[b_1,b_2,b_3\right]^\mathrm{T}\in\mathbb{C}^3$}
        
        \State $\varphi \gets  \angle b_2$
        
        \State \Return $\varphi$
        \EndFunction

    \end{algorithmic}
\end{algorithm}

\begin{algorithm}[t]
    \caption{ Used functions in Algorithm~\ref{algo:TMrecon}. Note that $\odot$, $\oslash$, and $^{\odot p}$ denote the element-wise multiplication, division, and $p$-th power, respectively;
        $\operatorname{exp}\{ \mathbf{A}\} \in \mathbb{C}^{m\times n}$,
        $\operatorname{abs}\{ \mathbf{A}\} \in \mathbb{R}^{m\times n}_+$,
        and $\operatorname{arg}\{ \mathbf{A}\} \in \left(-\pi,\pi\right]^{m\times n}$ are the element-wise
        exponential, modulus, and argument of a matrix $\mathbf{A}\in\mathbb{C}^{m\times n}$, respectively;  
        $\operatorname{diag}\{ \mathbf{B} \}\in\mathbb{C}^{m \times n}$ is a diagonal matrix made from the main diagonal of $\mathbf{B}\in\mathbb{C}^{m \times n}$,
        and $\mathbf{1}:=\left[1,1,\ldots,1\right]^\mathrm{T}\in\mathbb{R}^K_+$.} \label{algo:usedFunc}
    \begin{algorithmic}[1]
        \Function {SSMInitialization}{$\mathbf{X}, \mathbf{N}, \mathbf{F}$}
        \Comment{Speckle-correlation scattering matrix (SSM) \cite{lee2016exploiting}}
        
        \State Generate i.i.d. $\mathbf{T} \in \mathbb{C}^{M \times N}\sim \mathcal{N}\left(0,\frac{1}{2N}\right)+i\mathcal{N}\left(0,\frac{1}{2N}\right)$
        \State $\mathbf{N} \gets \mathbf{N}- \frac{1}{K} \mathbf{N}\mathbf{1}\mathbf{1}^\mathrm{T}$
        \Comment{Subtract the mean of the row vectors}
        \State $\mathbf{F} \gets \mathbf{F}- \frac{1}{K} \mathbf{F}\mathbf{1}\mathbf{1}^\mathrm{T}$
        \State $\mathbf{X} \gets \operatorname{diag}\{ \mathbf{X}\mathbf{X}^\dagger \}^{-1} \mathbf{X}$
        
        \While{$\mathbf{T}$ not converged}   
        \State $\mathbf{Y} \gets \mathbf{T}\mathbf{X}$
        \State $\mathbf{\widetilde{Y}} \gets \mathcal{F}\left\{\mathbf{Y}\right\}$
        \Comment{2D Fourier transform of each column vector of $\mathbf{Y}$}
        
        \State $\mathbf{T}_\mathrm{NF} \gets \left(\mathbf{N}\odot\mathbf{Y}\right) \mathbf{X}^\dagger$
        
        \State $\mathbf{T}_\mathrm{FF} \gets
        \mathcal{F}^{-1}\left\{ \left( \mathbf{F}\odot \mathbf{\widetilde{Y}} \right) \mathbf{X}^\dagger \right\} $
        \State $\mathbf{T} \gets \mathbf{T}_\mathrm{NF}+\mathbf{T}_\mathrm{FF}$
        \State $\mathbf{T} \gets \operatorname{diag}\{ \mathbf{T}\mathbf{T}^\dagger \}^{-\frac{1}{2}} \mathbf{T}$
        \Comment{Normalization of the row vectors}
        \EndWhile    
        \State \Return $\mathbf{T}$
        \EndFunction
        \Statex

        \Function {NearFieldResidual}{$\mathbf{Y}, \mathbf{N}$}
        \Comment{Smoothed amplitude flow \cite{luo2020phase}}
        \State $p \gets 4$
        \Comment{Smoothing parameter}        
        
        \State $\epsilon \gets 10^{-7}$
        \State $\mathbf{A}_\mathrm{NF}\gets\mathbf{N}^{\odot \frac{1}{2}}$
        
        \State $\mathbf{W}\gets \mathbf{Y} \oslash  \left( \mathbf{A}_\mathrm{NF} +\epsilon \right)$
        
        \State $\mathbf{W} \gets 
        \operatorname{abs}\left\{ \mathbf{W} \right\} $ 
        \State $\mathbf{J} \gets \frac{1}{K} \mathbf{W}\mathbf{1}\mathbf{1}^\mathrm{T} +\epsilon$ 
        
        \State $\mathbf{W} \gets \mathbf{W}\oslash\mathbf{J}$ 
        \State $\mathbf{U} \gets \left( \mathbf{W}^{\odot p} +\mathbf{J}^{\odot(-p)} \right)^{\odot \frac{1}{p}} +\epsilon$ 
        \State $\mathbf{U} \gets \left( 2^{\frac{1}{p}} - \mathbf{J} \odot \mathbf{U}\right) \odot \left( \mathbf{W} \oslash\mathbf{U}\right)^{\odot(p-1)}$ 
        \State $\mathbf{R}  \gets \mathbf{U}\odot \mathbf{A}_\mathrm{NF} \odot        
        \operatorname{exp}\{i 
        \operatorname{arg}\{ \mathbf{Y}  \} \} $ 
        \State \Return $\mathbf{R}$
        \EndFunction
        \Statex
        
        \Function {FarFieldResidual}{$\mathbf{Y}, \mathbf{F}$}
        \State $\mathbf{A}_\mathrm{FF}\gets\mathbf{F}^{\odot \frac{1}{2}}$
        \State $\mathbf{\widetilde{Y}} \gets \mathcal{F}\left\{\mathbf{Y}\right\}$
        \State $\mathbf{R} \gets 
        \mathcal{F}^{-1}\left\{
        \mathbf{A}_\mathrm{FF} \odot         \operatorname{exp}\{i 
        \operatorname{arg}\{ 
        \mathbf{\widetilde{Y}}
        \}
        \}
        \right\}-  \mathbf{Y}$
        \State \Return $\mathbf{R}$
        \EndFunction
    \end{algorithmic}
\end{algorithm}

\begin{algorithm}[t]
    \caption{ Phase-only SLM pattern optimization.
        Note that 
        $\operatorname{exp}\{ \mathbf{A}\}\in\mathbb{C}^{m\times n}$,
        $\operatorname{Im}\{ \mathbf{A}\}\in\mathbb{R}^{m\times n}$,
        and $\operatorname{arg}\{ \mathbf{A}\}\in\left(-\pi,\pi\right]^{m\times n}$ are the element-wise
        exponential, imaginary part, and argument of a matrix $\mathbf{A}\in\mathbb{C}^{m\times n}$, respectively.
    } \label{algo:SLMopt}
    \begin{algorithmic}[1]
        \State \textbf{Input:} 
        A normalized target input field $\mathbf{x} \in V$ and $\left\| \mathbf{x} \right\|_2=1$;
        the basis transform matrix $\mathbf{V} = \left[ \mathbf{v}_1, \mathbf{v}_2, \ldots, \mathbf{v}_{N_\mathrm{fib}} \right]$, where $\mathbf{v}_n$ is the $n$-th eigenvector of the global TM $\mathbf{T}^\dagger \mathbf{T}$;     
        the step size $\eta=1$; and the regularization parameter $\alpha_\mathbf{z}=0.05$.
        \Statex
        
        \Procedure {SLMPatternOptimization}{$\mathbf{x},\mathbf{V},\eta,\alpha_\mathbf{z}$} 
        \State $\bm{\theta} \gets \operatorname{arg}\{ \mathbf{x}\}$
        \State $\mathbf{m} \gets 0$        
        \State $\mu \gets 1$        
        \While{$\bm{\theta}$ not converged}   
        \State $\mathbf{m}_\mathrm{prev} \gets \mathbf{m}$        
        \State $\mu_\mathrm{prev} \gets \mu$
        
        \State $\mathbf{z} \gets \operatorname{exp}\{ i\bm{\theta}\}$   
        \State $\mathbf{z}_V \gets \mathbf{V}\mathbf{V}^\dagger\mathbf{z}$ 
        \Comment{$\mathbf{z}_V = \mathcal{P}_V \{ \mathbf{z} \}$}
        \State $\mathbf{r} \gets \left(
        \mathbf{z}_V-\left\| \mathbf{z}_V \right\|_2\mathbf{x}
        \right) -\alpha_\mathbf{z} \mathbf{z}_V$ 
        \State $\mathbf{g}_{\bm{\theta}} \gets 
        \operatorname{Im}\{\mathbf{V}\mathbf{V}^\dagger\mathbf{r} \odot \mathbf{z}^\ast
        \}$     
        \State $\mathbf{m} \gets \bm{\theta}- \eta \mathbf{g}_{\bm{\theta}}$
        \State $\mu \gets \frac{1+\sqrt{1+4\mu^2}}{2}$
        \State $\bm{\theta}\gets \mathbf{m}
        + \left( \frac{\mu_\mathrm{prev}-1}{\mu}   \right)
        \left( \mathbf{m}-\mathbf{m}_\mathrm{prev} \right)$
        \Comment{Nesterov's accelerated gradient \cite{nesterov1983method}}
        \EndWhile
        \State \Return $\bm{\theta}$
        \EndProcedure
    \end{algorithmic}
\end{algorithm}

\begin{algorithm}[t]
    \caption{ Output field optimization.
    } \label{algo:outputOpt}
    \begin{algorithmic}[1]
        \State \textbf{Input:} 
        A normalized target output field $\mathbf{b} \in \mathbb{C}^M$ and $\left\| \mathbf{b} \right\|_2=1$;
        The current transmission matrix $\mathbf{T}$, initialized by the measured global TM $\mathbf{T} \in \mathbb{C}^{M \times N}$;
        $\mathbf{P}_{o,e}\in\left\{0,1\right\}^{\frac{M}{2}\times M}$ that extracts the ordinary or extraordinary polarizations of the beam displacer from the near- or far-field images;
        the inner iteration number $J=100$;
        the local domain size determination parameter $\varepsilon=0.95$;
        the oversampling ratio for the local TM reconstruction $\gamma_\mathrm{loc}=7$;
        the step size $\eta = 0.1$;
        and the stopping criteria.    
        \Statex
        \Procedure {OutputFieldOptimization}{$\mathbf{b},\mathbf{T},\mathbf{P}_o, \mathbf{P}_e,J,\varepsilon,\gamma_\mathrm{loc},\eta$} 
        \State $\mathbf{x} \gets \mathbf{T}^+\mathbf{b}$  
        \Comment{Initialization using the pseudoinverse of $\mathbf{T}$}
        \State $\mathbf{x} \gets \mathbf{x}/\|\mathbf{x}\|_2$
        
        \While{true}   
        \State $C^\mathrm{max}_\mathbf{b}\gets 0$
        \For{$j = 1$ to $J$}   
        \State $\mathbf{n},\;\mathbf{f},\;\mathbf{n}^{\scriptscriptstyle\lambda/4}, \;\mathbf{f}^{\scriptscriptstyle\lambda/4} \gets$ \Call{DataAcquisition}{$\mathbf{x}$}      
        \Comment{In both linear and circular polarizations}
        
        \State $\mathbf{y} \gets$ \Call{FieldEstimation}{$\mathbf{n}, \mathbf{f}, \mathbf{n}^{\scriptscriptstyle\lambda/4}, \mathbf{f}^{\scriptscriptstyle\lambda/4}, \mathbf{P}_o, \mathbf{P}_e, \mathbf{b}$}      
        \Comment{See Algorithm~\ref{algo:GSalgo}}
        \State $C_{\mathbf{b}} \gets \left| \mathbf{y}^\dagger \mathbf{b} \right| $        
        
        \If{the stopping criteria are met}
        \State \Return $\mathbf{x},\mathbf{y},\mathbf{T}$
        \EndIf
        \If{$C_{\mathbf{b}} > C^\mathrm{max}_\mathbf{b}$}
        \State $C^\mathrm{max}_\mathbf{b} \gets C_{\mathbf{b}}$
        \State $\mathbf{x}_0 \gets \mathbf{x}$
        \Comment{Save $\mathbf{x}$ that provides the highest correlation}
        \EndIf
        \State $\mathbf{r} \gets \mathbf{b}-\mathbf{y}$
        \State $\mathbf{x} \gets\mathbf{x} + \eta \mathbf{T}^\dagger \mathbf{r}$
        \State $\mathbf{x} \gets \mathbf{x}/\|\mathbf{x}\|_2$
        \EndFor
        
        \State $\rho \gets$ \Call{SetLocalDomainSize}{$\mathbf{x}_0,\mathbf{b},\varepsilon$}  
        \Comment{Eq.~\ref{eq:optLocalSize}, \textit{fminbnd} function of MATLAB}
        
        \State $\mathbf{T} \gets$ \Call{LocalTMMeasurement}{$\mathbf{x}_0,\rho,\gamma_\mathrm{loc}$}  
        \Comment{See Methods}
        \State $\mathbf{x} \gets \mathbf{x}_0$
        \Comment{Initialize the next stage by the best input in the previous stage}
        \EndWhile
        \EndProcedure
    \end{algorithmic}
\end{algorithm}

\begin{algorithm}[t]
    \caption{ Modified Gerchberg--Saxton algorithm used in Algorithm~\ref{algo:outputOpt}.
        Note that $\odot$, $\oslash$, and $^{\odot p}$ denote the element-wise multiplication, division, and $p$-th power, respectively;
        $\operatorname{exp}\{ \mathbf{A}\}\in\mathbb{C}^{m\times n}$
        and $\operatorname{arg}\{ \mathbf{A}\}\in\left(-\pi,\pi\right]^{m\times n}$ are the element-wise
        exponential and argument of a matrix $\mathbf{A}\in\mathbb{C}^{m\times n}$, respectively.
    } \label{algo:GSalgo}
    \begin{algorithmic}[1]
        
        \Statex
        \Function {FieldEstimation}{$\mathbf{n}, \mathbf{f}, \mathbf{n}^{\scriptscriptstyle\lambda/4}, \mathbf{f}^{\scriptscriptstyle\lambda/4}, \mathbf{P}_o, \mathbf{P}_e, \mathbf{b}$}
        \Comment{Modified Gerchberg--Saxton algorithm}
        \State $\mathbf{n}_\mathrm{H,V}\gets
        \mathbf{P}_{o,e}\mathbf{n}$
        \State $\mathbf{n}_\mathrm{R,L} \gets
        \mathbf{P}_{o,e}\mathbf{n}^{\scriptscriptstyle\lambda/4}$
        \State $\mathbf{f}_\mathrm{H,V}\gets
        \mathbf{P}_{o,e}\mathbf{f}$    
        \State $\mathbf{f}_\mathrm{R,L} \gets
        \mathbf{P}_{o,e}\mathbf{f}^{\scriptscriptstyle\lambda/4}$    
        \State $\mathbf{b}_\mathrm{H,V} \gets \mathbf{P}_{o,e}\mathbf{b}$
        \State $\mathbf{b}_\mathrm{R,L} \gets
        \frac{1}{\sqrt{2}} \left( \mathbf{b}_\mathrm{H} \mp i \mathbf{b}_\mathrm{V}\right)$

        \State $\mathbf{A}_\mathrm{NF} \gets \left[ 
        \mathbf{n}_\mathrm{H}^{\odot \frac{1}{2}},
        \mathbf{n}_\mathrm{V}^{\odot \frac{1}{2}},
        \mathbf{n}_\mathrm{R}^{\odot \frac{1}{2}},
        \mathbf{n}_\mathrm{L}^{\odot \frac{1}{2}}
        \right]$ 
        \State $\mathbf{A}_\mathrm{FF} \gets \left[ 
        \mathbf{f}_\mathrm{H}^{\odot \frac{1}{2}},
        \mathbf{f}_\mathrm{V}^{\odot \frac{1}{2}},
        \mathbf{f}_\mathrm{R}^{\odot \frac{1}{2}},
        \mathbf{f}_\mathrm{L}^{\odot \frac{1}{2}}
        \right]$   
        \Comment{$\mathbf{A}_\mathrm{NF,FF} \in\mathbb{R}^{\frac{M}{2}\times 4}_+$}

        \State $\mathbf{B}_\mathrm{HVRL} \gets \left[
        \mathbf{b}_\mathrm{H},
        \mathbf{b}_\mathrm{V},
        \mathbf{b}_\mathrm{R},
        \mathbf{b}_\mathrm{L}
        \right]$   
        \Comment{$\mathbf{B}_\mathrm{HVRL} \in\mathbb{C}^{\frac{M}{2}\times 4}$}
        
        \State $\mathbf{Q}= \frac{1}{\sqrt{2}}\left[ 1, -i; 1, +i \right]$
        \State $\mathbf{W} = \frac{1}{2}\left[ \mathbf{I}_2, \mathbf{Q}^\dagger; \mathbf{Q},\mathbf{I}_2 \right]$
        \Comment{$\mathbf{I}_2$ is the 2D identity matrix, $\mathbf{W} \in\mathbb{C}^{4\times 4}$}    
        \State $\mathbf{Y}_\mathrm{HVRL}  \gets \mathbf{A}_\mathrm{NF} \odot  
        \operatorname{exp}\{i \operatorname{arg}\{ \mathbf{B}_\mathrm{HVRL} \} \} $ 
        \Comment{Initialization with the target phase}    
        \State $\mathbf{M} \gets \mathbf{Y}_\mathrm{HVRL}$
        \State $\mu \gets 1$  
        \While{$\mathbf{Y}_\mathrm{HVRL}$ not converged}   
        \State $\mathbf{M}_\mathrm{prev} \gets \mathbf{M}$
        \State $\mu_\mathrm{prev} \gets \mu$
        
        \State $\mathbf{\widetilde{Y}}_\mathrm{HVRL} \gets \mathcal{F}\{\mathbf{Y}_\mathrm{HVRL}\}$
        \Comment{2D Fourier transform of each column vector of $\mathbf{Y}_\mathrm{HVRL}$}
        \State $\mathbf{R} \gets \mathbf{\widetilde{Y}}_\mathrm{HVRL} 
        -\mathbf{A}_\mathrm{FF}\odot \operatorname{exp}\{i \operatorname{arg}\{ 
        \mathbf{\widetilde{Y}}_\mathrm{HVRL}  \} \}$              
        
        \State $\mathbf{M} \gets \mathbf{Y}_\mathrm{HVRL} - \mathcal{F}^{-1}\left\{ \mathbf{R} \right\}$
        \State $\mu \gets \frac{1+\sqrt{1+4\mu_\mathrm{prev}^2}}{2}$
        \State $\mathbf{Y}_\mathrm{HVRL} \gets \mathbf{M}
        + \left( \frac{\mu_\mathrm{prev}-1}{\mu}   \right)
        \left( \mathbf{M}-\mathbf{M}_\mathrm{prev} \right)$  
        \Comment{Nesterov's accelerated gradient \cite{nesterov1983method}}
        \State $\mathbf{Y}_\mathrm{HVRL}  \gets \mathbf{A}_\mathrm{NF} \odot  
        \operatorname{exp}\{i \operatorname{arg}\{ \mathbf{Y}_\mathrm{HVRL} \} \} $ 
        \State $\mathbf{Y}_\mathrm{HVRL} \gets \mathbf{Y}_\mathrm{HVRL}\mathbf{W}$ 
        \Comment{Induce crosstalk between polarizations}
        \EndWhile    
        \State $\left[ 
        \mathbf{y}_\mathrm{H}, \mathbf{y}_\mathrm{V}, \mathbf{y}_\mathrm{R},\mathbf{y}_\mathrm{L}  \right]
        \gets \mathbf{Y}_\mathrm{HVRL}$
        \Comment{$\mathbf{y}_\mathrm{H}, \mathbf{y}_\mathrm{V}, \mathbf{y}_\mathrm{R}, \mathbf{y}_\mathrm{L} \in \mathbb{C}^{M/2}$}
                
        \State $\mathbf{y} \gets 
        \mathbf{P}_o^\mathrm{T} \mathbf{y}_\mathrm{H} + \mathbf{P}_e^\mathrm{T} \mathbf{y}_\mathrm{V}$
        \Comment{$\mathbf{y} \in \mathbb{C}^M$}
        \State $\mathbf{y}\gets\mathbf{y}/\left\| \mathbf{y}\right\|_2$
        \State $\mathbf{y} \gets \mathbf{y} \operatorname{exp}\{i \angle(\mathbf{y}^\dagger \mathbf{b})\}$          
        \Comment{Global phase matching}
        
        \State \Return $\mathbf{y}$
        \EndFunction
        
    \end{algorithmic}
\end{algorithm}



\end{document}